\documentclass[default,sn-mathphys-ay,iicol]{sn-jnl}

\usepackage{lmodern}
\usepackage{caption}
\usepackage{subcaption}
\usepackage{textgreek}
\usepackage{tablefootnote} 
\usepackage{tabularx}
\usepackage{microtype}
\usepackage{hyperref}
\usepackage{multirow}
\usepackage{makecell}
\usepackage{amsmath}
\usepackage{mathtools}
\usepackage{empheq}
\usepackage{esvect}
\usepackage{booktabs}
\usepackage{graphicx} 
\usepackage{siunitx}
\usepackage[section]{placeins}
\usepackage{nomencl}
\usepackage{caption,subcaption} 
\usepackage{bm}
\usepackage[normalem]{ulem} 
\usepackage{standalone} 

\begin{document}

\title[Transonic buffet on OAT15A wing subjected to 2DOF]{Experiments on an airfoil with adjustable pitch and heave under transonic buffet leading to lock-in coupling}


\author*[1]{\fnm{Cédric} \sur{Thémiot}}\email{cedric.themiot@onera.fr}
\author[1]{\fnm{Vincent} \sur{Brion}}\email{vincent.brion@onera.fr}
\author[1]{\fnm{Arnaud} \sur{Lepage}}\email{arnaud.lepage@onera.fr}

\affil*[1]{\orgdiv{DAAA}, \orgname{ONERA}, \orgaddress{\street{Institut Polytechnique de Paris}, \postcode{92320}, \city{Chatillon}, \country{France}}}

\abstract{
This paper presents aeroelastic experiments conducted in the ONERA S3Ch transonic wind tunnel using a supercritical OAT15A airfoil mounted on a rig allowing independent control of pitch and heave degrees of freedom (DOF). 
The study investigates fluid-structure interaction (FSI) mechanisms and lock-in phenomena between transonic buffet and structural modes. 
Aerodynamic and structural responses were monitored using steady and unsteady pressure sensors, high-speed stereoscopic Particle Image Velocimetry (PIV), accelerometers, and optical sensors. 
Four configurations were tested: rigid (0 DOF), pitch-only (1 DOF pitch), heave-only (1 DOF heave), and combined pitch-heave (2 DOF). 
The test campaign spanned selected Mach numbers, angles of attack, and structural natural frequencies for each configuration. 
Results show that buffet-induced lock-in occurs abruptly when the Mach number increases, but only when the pitch DOF is enabled, in both the 1-DOF pitch or 2-DOF configuration. 
These findings establish that pitch-buffet frequency proximity is a necessary condition for lock-in.
In the 2-DOF configuration, continuous Mach number excursions revealed a three-way coupling involving the buffet, pitch, and bending modes, with the bending mode modulating the coupled dynamics. 
This coupling generates large-amplitude limit-cycle oscillations (LCO) within seconds, persisting over a narrow Mach range before buffet cessation. 
}





\maketitle

\clearpage

\section{Introduction}\label{sec:intro}

In the transonic flow regime, interactions between shock waves and separated shear layers can give rise to large-amplitude self-sustained instabilities known as transonic shock buffet.
These shock wave oscillations were observed several decades ago (\cite{hilton1947photographs,mcdevitt1985static}), and resulted in the categorization of two types of shock buffet.
Type I usually occurs on biconvex airfoils with no Angle of Attack (AoA) and includes shock oscillations on both sides of the wing.
Type II involves shock oscillations on the suction side of supercritical airfoils, for certain Mach numbers and AoA.
The shock buffet phenomenon is governed by upper surface flow separation, yet the root cause and physics governing these shock oscillations remain a subject of investigation.
\cite{pearcey1955some} and \cite{pearcey1968interaction} established fundamental separation models (shock-induced bubble or trailing-edge separation) and linked buffet onset to the extension of the separation bubble to the trailing edge.
While this theory was long the consensus, recent studies have reassessed it (\cite{crouch2009origin,nitzsche2009numerical}), noting shock motion inversion as a preliminary sign of onset (\cite{accorinti2022experimental}).
\cite{tijdeman1977investigations} further detailed shock motion dynamics on NACA 64A006 airfoils equipped with oscillating flaps, classifying them into three categories. Comprehensive reviews by \cite{lee2001self} and \cite{giannelis2017review} consolidate these findings and outline remaining challenges.

Transonic shock buffet on fixed wings was thoroughly investigated by~\cite{jacquin2009experimental,jacquin2016testing} using a supercritical OAT15A airfoil in the ONERA S3Ch wind tunnel.
The database established buffet onset boundaries and noted that the natural buffet frequency increases with Mach number and AoA.
This campaign serves as a baseline for the present study.

Transonic shock buffet can trigger aeroelastic Fluid-Structure Interactions (FSI) when coupled with structural Degrees Of Freedom (DOF), potentially leading to a lock-in phenomenon.
Raveh and Dowell conducted an in-depth numerical investigation on a NACA 0012 airfoil under forced heave oscillations~(\cite{raveh2009numerical,raveh2011frequency}).
For low-amplitude excitations and/or when the excitation frequency is sufficiently remote from the natural buffet frequency, the aerodynamic response exhibits two dominant frequency components, corresponding to the buffet and the forced airfoil motion.
However, as the excitation amplitude increases and its frequency approaches the buffet frequency, a lock-in phenomenon emerges: the response at the buffet frequency diminishes and is replaced by a dominant response at the motion frequency.
The frequency range over which this lock-in occurs expands with increasing oscillation amplitude.
Notably, lock-in can occur for excitation frequencies both below and above the buffet frequency, though the resulting aerodynamic behavior differs.
When the excitation frequency lies below the buffet frequency, the shock exhibits dynamics akin to classic transonic buffet, characterized by strong oscillations and significant variations in shock intensity.
In contrast, when the excitation frequency exceeds the buffet frequency, the shock response becomes quasi-harmonic, with reduced oscillation amplitude and weaker variations in shock strength.

\cite{giannelis2017investigation} obtained comparable results in their numerical study, which was based on the experimental campaign conducted by~\cite{jacquin2009experimental} at ONERA.
Their work involved applying forced pitch oscillations to a supercritical OAT15A airfoil, thereby extending the experimental findings of Raveh and Dowell on a symmetrical NACA 0012 profile to the broader context of supercritical airfoils.
Following this numerical study, \cite{hartmann2013coupled} performed a wind tunnel test campaign on a supercritical DRA 2303 airfoil in buffet conditions.
Measurements included a fixed airfoil configuration, flow-induced motion with a flexible test setup having one heave and pitch stiffness frequency, and forced heave and pitch oscillations in the vicinity of the buffet frequency.
Results regarding the forced oscillation configuration confirmed the findings of Raveh and Dowell, displaying lock-in of the flow frequency to the forced motion frequency slightly below, above and at the buffet frequency.
Without forced oscillations, observations at the shock location revealed an increase in fluctuation levels at the buffet frequency for the fixed airfoil, which was not visible anymore in the flow-induced motion flexible configuration.
Also, the amplitude of shock oscillation decreased, which is attributed to an energy transfer from the fluid into the structure.
No change in the frequency of oscillation or lock-in phenomenon on either the heave or pitch natural frequencies of the setup was observed, contrary to the effects of forced oscillations to the experiment at the vicinity of buffet frequency.
The pronounced difference between the structural (pitch and heave) frequencies and the buffet frequency (about 1/3 and 1/2 respectively) can account for the absence of a lock-in phenomenon.
Therefore, the test campaign presented in the present article will be aimed at investigating FSI on a supercritical OAT15A profile, in a system with free structural degrees of freedom, in which the pitch and heave frequencies can be tuned to approach the buffet frequency.

The literature includes numerical simulations focused on similar test configurations.
\cite{raveh2014aeroelastic} performed Navier-Stokes simulations to determine the responses of a single-DOF pitch and a two-DOF pitch-and-heave NACA 0012 airfoil subjected to transonic buffet.
They observed that when the natural frequency in pitch was lower than the buffet frequency, lock-in did not occur. The system responded harmonically at the buffet frequency, with the shock acting as an external excitation.
In contrast, when the pitch natural frequency exceeded the buffet frequency, the system gradually transitioned to a large-amplitude Limit-Cycle Oscillations (LCO) characterized by the pitch structural frequency alone.
Raveh and Dowell also examined the influence of structural damping in the single-DOF pitch configuration.
Their results show that any positive damping rapidly suppresses the coupling: even minimal values ($\varsigma = 0.05\%$) delay the growth of the pitching response and reduce the LCO amplitude, and a damping of only ($\varsigma = 0.5\%$) is sufficient to suppress lock-in entirely.
In a coupled pitch-heave configuration with natural frequencies slightly offset from the buffet frequency ($f_{\text{pitch}}  = 1.1 f_{\text{buffet}}$, $f_{\text{heave}} = 0.9 f_{\text{buffet}}$), the system exhibited large-amplitude pitch LCO, with aerodynamic and structural responses locked-in at the pitch frequency.
A linear flutter analysis confirmed that this instability is not flutter-induced.
When the natural frequencies were reversed ($f_{\text{pitch}}  = 0.9 f_{\text{buffet}}$, $f_{\text{heave}} = 1.1 f_{\text{buffet}}$), the response was limited to low-amplitude oscillations at the buffet frequency, consistent with prior observations~(\cite{raveh2009numerical,raveh2011frequency,iovnovich2012transonic}).

\cite{giannelis2016aeroelastic} extended these findings for an elastically suspended supercritical OAT15A airfoil.
Simulations were performed for several mass ratios and structural frequencies (below and above the shock buffet frequency).
Results were not very dependent on the mass ratio, however the relative structural and aerodynamic frequencies were critical. Lock-in with the structural motion frequency (pitch or heave) occurred when the structural frequency was above and below the shock buffet frequency, respectively.
This behavior aligns with phase analyses reported earlier for airfoils undergoing forced oscillations or airfoils suspended in pitch or in combined pitch-heave configurations (\cite{raveh2009numerical,raveh2011frequency,raveh2014aeroelastic,giannelis2017investigation}).
\cite{giannelis2016dynamic} also demonstrated that, in a system with a single pitch DOF, the occurrence of lock-in was primarily governed by the ratio of the pitch natural frequency to the shock-buffet frequency.
The lock-in region was observed within the interval $1 < f_{\text{pitch}}/f_{\text{buffet}} < 1.8$, with the largest oscillation amplitudes occurring near a ratio of 1.5.

Recently, an experimental test campaign was performed on an OAT15A supercritical airfoil (chord length $c = 152mm$, span $s = 298mm$, aspect ratio $AR = 2$) in transonic conditions and varying angles of attack in the Trisonic Wind Tunnel of the Bundeswehr University in Munich (TWM), with one DOF (pitch) (\cite{accorinti2022experimental, kokmanian2022investigating, scharnowski2022shock, korthauer2023effect, korthauer2023experimental}). This test campaign and the one presented in this paper complement each other, as they were both part of European Union Horizon 2020 "HOMER" Project (Holistic Optical Metrology for Aero-Elastic Research).

\cite{accorinti2022experimental} investigated the transonic shock buffet phenomenon on the rigid configuration, without the pitch DOF, at varying transonic Mach numbers ($0.71 < M_0 < 0.78$) and AoA ($2.5^\circ < \alpha < 6.5^\circ$), focusing on buffet boundaries, comparing the conditions leading to the inversion of shock motion and those displaying buffet onset.
Researchers showed that shock motion inversion is a necessary but not sufficient condition for buffet onset.
While trends aligned with existing literature, this study found lower buffet amplitudes and higher angle of attack thresholds for buffet onset.
It was observed that $\alpha_0$ needed to be increased by 0.3-0.4° after the shock motion inversion started to occur in order to observe buffet onset.
Additionally, the influence of experimental conditions such as boundary-layer suction at the vertical walls and the gap allowing leakage flow at the side windows was investigated, revealing a slight impact on buffet frequencies and amplitudes, but negligible effect on buffet boundaries.

\cite{kokmanian2022investigating} investigated the same rigid configuration using Particle Image Velocimetry (PIV) to analyze the flow features.
Consistent with observations from \cite{jacquin2009experimental}, the study found that the shock most upstream position coincides with the formation of an extensive separation region, highlighting a recurring pattern in transonic flow behavior.
Spectral analysis confirmed a direct correlation between increasing transonic Mach numbers ($M_0$) and AoA ($\alpha_0$) and an increase in dominant frequencies of the shock motion, an observation consistent with surface pressure measurements, validating earlier findings by \cite{giannelis2018influence} and \cite{accorinti2022experimental}.
Based on PIV measurements, the study further characterized the downstream evolution of coherent structures shed by the shock.
Their convective velocity increased notably with streamwise position, from approximately 50 m/s at $x/c = 0.6$ to nearly 150 m/s at $x/c = 1.1$, indicating a clear acceleration of the structures as they propagate away from the shock.

\cite{scharnowski2022shock} explored the impact of introducing a spring-enabled flexible pitch DOF to the OAT15A airfoil model experiment, investigating the interaction between transonic flow and induced pitching motion.
One pitch frequency was tuned by the test setup, in the vicinity of the buffet frequency, at $M_0 = 0.74$ and a mean AoA $\alpha_0 = 5.8^\circ$.
Results showed the presence of a periodic pitch motion with an amplitude of $\pm 0.9^\circ$.
The model structural pitch motion synchronized with the natural buffet frequency, locking into a dominant frequency of 115 Hz, slightly above the buffet frequency of the rigid model and the structural pitch natural frequency without flow.
High-temporal-resolution PIV measurements captured the evolution of the buffet cycle, showing strong variations in shock position, compression shape, and boundary layer state downstream of the shock.
Notably, the pitch DOF significantly amplified flow field dynamics compared to a rigid model, leading to substantial changes in wake velocity deficit, drag, and lift during the buffet cycle. A nonlinear relationship between AoA and drag was found to limit oscillation amplitude, preventing model destruction.

\cite{korthauer2023effect,korthauer2023experimental} comprehensively explored the effects of varying structural (pitch frequencies) and fluid (AoA, Mach number) parameters on transonic buffet phenomenology in the elastically suspended wing configuration.
Key findings revealed that increasing Mach numbers at constant structural settings linearly decreased the buffet onset AoA, consistent with results observed on a NACA 0012 airfoil by \cite{mcdevitt1985static}, while changes in pitch eigenfrequency substantially impacted this boundary, with the higher tested pitch frequencies leading to premature buffet onset across all Mach numbers.
Results also displayed distinct fluid-structure interaction behaviors, depending on the relative position between pitch and buffet frequency.
Pitch to buffet frequency ratios $f_{\text{pitch}}/f_{\text{buffet}}$ of 0.9 and 1 (pitch below or matching the buffet frequency) were found to be associated with fluid-mode flutter behavior.
However, a higher ratio of 1.2, with pitch above buffet frequency, transitioned into structural-mode flutter.
In this configuration, once a frequency lock-in case was established, the role of the structural mode was predominant.
These lock-in interactions were also characterized by high yet limited (LCO type) pitch amplitudes.
The mass ratio played a more significant role in the interaction than expected, influencing the amplitude of pitch motion and introducing prematurity of buffeting onset at lower mass ratios.

Within this wide context, ONERA conducted a wind tunnel test campaign to experimentally investigate buffet occurrence and potential FSI (\cite{themiot2023wind,sicard2025experimental}).
The insights seen in numerical predictions above played a key role in guiding the design of the experimental setup and the selection of target frequency ranges for the heave and pitch DOFs.
Heave frequencies were chosen to lie below the shock buffet frequency, while pitch frequencies were set above it, with the objective of capturing large-amplitude oscillations and investigating the lock-in phenomenon.

This test campaign shares aerodynamic similarities with the previous campaign conducted by \cite{jacquin2009experimental}, utilizing a similar OAT15A supercritical airfoil (this time with a chord length $c = 250 mm$) and the ONERA S3Ch transonic wind tunnel facility.
However, the structural boundary conditions differ in this campaign, which includes two DOFs in pitch and heave.
Also, the wind tunnel walls were different, featuring a window on the suction side of the model, to allow PIV measurements.
The experimental setup enables independent adjustment of each DOF, allowing the test matrix to systematically investigate configurations with 0 DOF (rigid), 1 DOF (varying either pitch or heave frequency), and 2 DOF (combined pitch and heave).

The present article is aimed at describing the experimental test campaign, its design, preparation, instrumentation, and showcase the observed results for the different tested configurations.

\section{Theoretical background}

\subsection{Modeling of the test setup}

\begin{figure}
    \centering
    \includegraphics[width=\columnwidth]{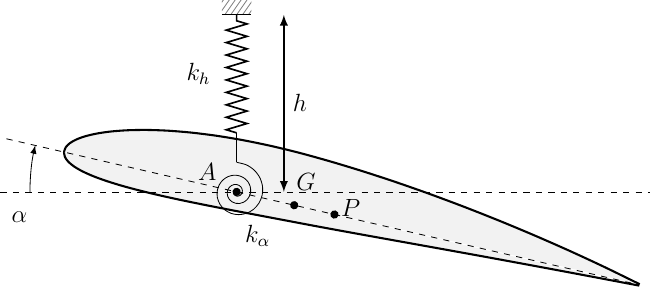}
    \caption{Theoretical drawing of the airfoil subjected to shock buffet, with pitch and heave DOF}
    \label{fig:theory_schematics}
\end{figure}

The aeroelastic problem is formulated using the configuration shown in Fig.~\ref{fig:theory_schematics}.  
The airfoil is mounted on two springs, allowing vertical (heave) and rotational (pitch) motions. 
Heave stiffness and vertical displacements are noted $k_h$ and $h$, while pitch stiffness and angular displacements are noted $k_\alpha$ and $\alpha$.
In the experiment, the pitch axis of rotation $A$ is located at $x_A/c = 0.30$, the center of gravity $G$ at $x_G/c \approx 0.33$, and the center of pressure $P$ at $x_P/c \approx 0.39$ (at $M=0.73$), with $c$ refering to the chord of the airfoil.

\subsection{Aerodynamic parameters}
\label{sec:aero_conditions}

The key aerodynamic input for the experiment is the expected buffet frequency range.
Since transonic buffet frequency depends primarily on the Mach number, angle of attack, and the airfoil's characteristic length scale, the structural design parameters -- particularly the target heave and pitch natural frequencies -- were then selected to lie close to this range.

To estimate the buffet frequency range of the present model, the experimental results of \cite{jacquin2009experimental} were used as a reference.  
Because their OAT15A model had a smaller chord ($c=230$ mm versus 250 mm here), the expected frequencies were rescaled by assuming a constant Strouhal number,
$St = \frac{f c}{U_0}$,
where $f$ is the characteristic buffet frequency, $c$ the chord, and $U_0$ the freestream velocity.
\nomenclature{$St$}{Strouhal number}
\nomenclature{$c$}{Airfoil chord length [m]}
\nomenclature{$b$}{Airfoil span length [m]}
\nomenclature{$U_0$}{Freestream velocity [m\,s$^{-1}$]}
This assumption preserves dynamic similarity between the two configurations.  
\cite{jacquin2009experimental} reported buffet frequencies between 65 and 77~Hz for Mach numbers between 0.72 and 0.74.  
Applying the chord correction yields a predicted buffet range of approximately $60 \text{Hz} < f_{\text{predicted buffet}} < 71 \text{Hz}$ for the present model under similar flow conditions.  
The heave and pitch frequencies of the test rig were therefore designed around this target range.

Also, the mean surface pressure coefficient $K_p$ will be discussed in upcoming figures, and is defined as $K_p = \frac{P - P_0}{\frac{1}{2} \rho_0 U_0^2}$, 
in which $P$ corresponds to the static pressure at the level of the sensor, $P_0$ to the free stream atmospheric pressure and $\rho_0$ to the free stream air density.
\nomenclature{$K_p$}{Mean surface pressure coefficient}
\nomenclature{$P$}{Static pressure [Pa]}
\nomenclature{$P_0$}{Free stream stagnation pressure [Pa]}
\nomenclature{$\rho_0$}{Free stream air density [kg m$^{-3}$]}

\subsection{Structural parameters}
\label{sec:struct_params}

The main structural parameters governing the aeroelastic response are the natural frequencies, mass and inertia ratios, and the relative positions of the center of gravity, center of pressure, and pitch rotation axis.  
Because each DOF is associated with a dedicated mechanical subassembly, the effective masses in motion differ between heave and pitch ($m_{heave}~\approx~47$~kg and $m_{pitch}~\approx~28$~kg).

\section{Experimental test setup}
\label{sec:expe_setup}

\subsection{Wind tunnel features}

The present study was carried out in the continuous closed-circuit transonic S3Ch wind tunnel of the ONERA - Meudon Center.
This facility is powered by a 3500 kW two-stage motor-ventilator group. The Mach number domain can extend from 0.3 to 1.2, the stagnation pressure being the atmospheric pressure, and the stagnation temperature ranging from \qtyrange{290}{310}{\kelvin}.
The flow is controlled either by a downstream sonic throat that allows setting the Mach number of the test section, or by rotating the angle of attack of the engine blades, keeping rotation speed constant\footnote{The second option generates a slight degradation in Mach number uncertainty, which was acceptable and allowed to perform some stabilized Mach number values and Mach number transient sweeps that could not have been achieved with the first one.}.

\subsection{Test rig with pitch and heave DOF stiffness adjustment}
\begin{figure}[h]
     \centering
        \includegraphics[width=\columnwidth]{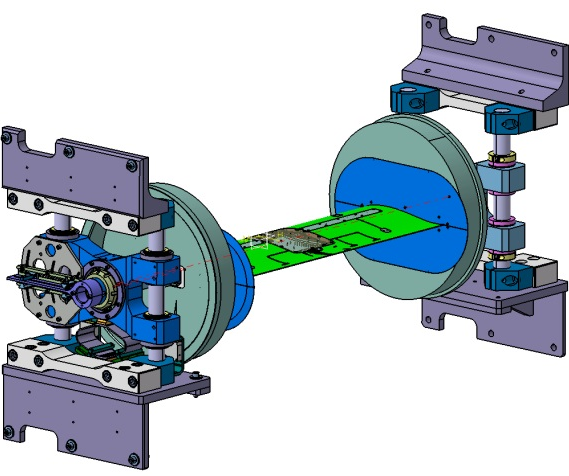}
        \caption{CAD view of the test setup}
        \label{fig:CAD_test_setup}
\end{figure}

The test setup is based on an existing setup that was designed for previous test campaigns investigating the gust load, conducted at ONERA S3Ch (\cite{lepage2015complete}), adapted to meet this test technical specifications.
A CAD view of the setup is visible in Fig.~\ref{fig:CAD_test_setup}.

\begin{figure}[h]
         \centering
         \includegraphics[width=\columnwidth]{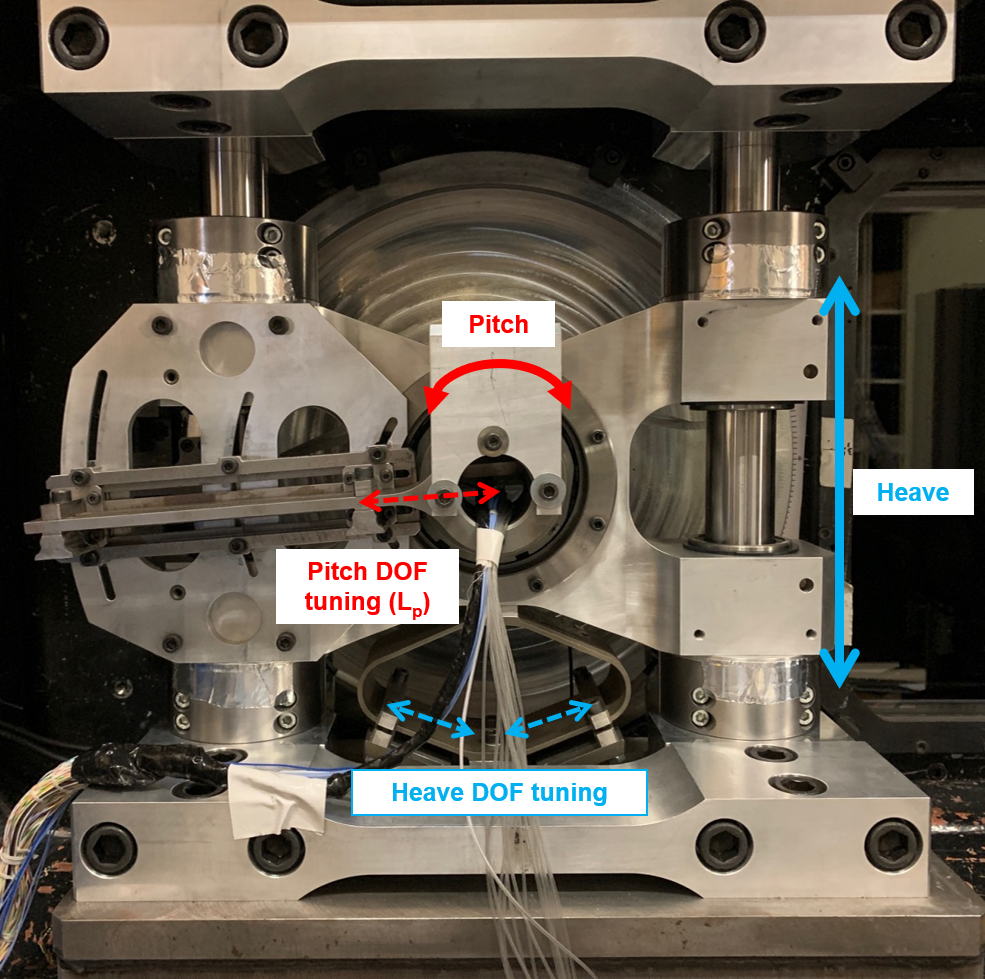}
        \caption{Description of DOF adjustments (pitch: red, heave: blue). Displayed: O DOF configuration (pitch and heave DOF are at their stiffest setting, and additional pitch and heave DOF blocking parts are installed). $\alpha_0 = 3.5^{\circ}$.}
        \label{fig:DOF_test_setup}
\end{figure}
    
One side of the actual test setup is visible in Fig.~\ref{fig:DOF_test_setup}.
It allows the adjustment of the static targeted AoA ($\alpha_0$).
The stiffness of the pitch and heave (also often called plunge) DOF can both be adjusted by sliding parts that change the free lengths of the spring blades.
For the pitch DOF, sliding the tuning part to the right reduces the length parameter $L_p$ to increase the pitch frequency, providing a continuous tuning of the stiffness.
For the heave DOF, two tuning parts can be moved along the bottom half of a diamond-shaped spring part.
Several stiffness settings were available, and were qualitatively named 0 (tuning parts removed, most flexible), 1 (tuning parts in lowest position), 2 (middle position) and 3 (high position, stiffest).
Additional parts could be added to bypass and block the pitch and/or heave DOF as rigidly as possible, as in Fig.~\ref{fig:DOF_test_setup}.

\nomenclature{$L_p$}{Length parameter governing the pitch DOF stiffness (cf Fig.~\ref{fig:DOF_test_setup})}

As reported in the introduction, it appeared that aeroelastic instability and potential lock-in were to be expected rather when the pitch frequency was above shock buffet frequency, and when heave frequency was below shock buffet frequency. The rig parts (especially the DOF springs) were designed via Finite Element Analysis to meet these requirements, and the test matrix was chosen to investigate these conditions.
The mass ratios for the pitch and heave motions are significantly larger than those investigated in \cite{giannelis2016dynamic}, yet, the latter's findings indicated that mass ratio was not the driving parameter influencing the onset of lock-in phenomenology.

The sealing between the interior and exterior of the test section surrounding the model was specifically designed to permit unrestricted motion of the two degrees of freedom (DOF), even for minimal energy transfer due to fluid-structure interactions.
This feature was considered essential, as \cite{raveh2014aeroelastic} demonstrated that even very small levels of damping ($\varsigma = 0.5\%$) can suppress the lock-in phenomenon under certain conditions.
The design employed a wall-mounted labyrinth seal to eliminate frictional effects, complemented by splitter plates to prevent residual leakage from disturbing the flow over the wing.
The effectiveness of the solution was verified through an oil flow visualization test in the first phase of the wind tunnel test campaign.

\subsection{Model and instrumentation}

\begin{figure*}
    \centering
    \includegraphics[width=14cm]{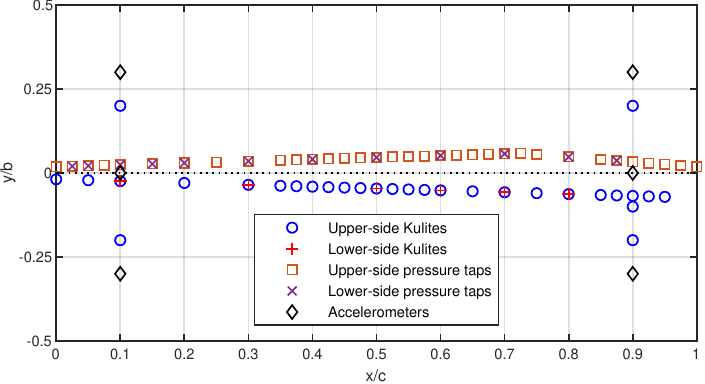} 
    \caption{Top view of instrumentation distribution of the test model}
    \label{fig:Instru}
\end{figure*}

The model consists of an OAT15A airfoil with a relative thickness of 12.3\%, a chord length $c~=~250$~mm, an effective span $b~=~752$~mm between the splitter plates (resulting in an aspect ratio $AR~=~3$), and a trailing-edge thickness of 0.5\% of the chord.
The representativeness of this aspect ratio is supported by \cite{accorinti2023characterization}, who demonstrated on an OAT15A airfoil with $AR~=~2$ that the amplitude of shock oscillations becomes nearly uniform along most of the span once buffet is established.
Boundary-layer tripping was applied at $x/c~=~7\%$ using CAD-cut strips of 89/102~µm height, chosen to match the local boundary-layer thickness on both the suction and pressure sides.

Fig.~\ref{fig:Instru} shows a top view of the instrumentation distribution of the test model, normalized with chord and span ($c$ and $b$ respectively).
The airfoil was equipped with six internal accelerometers (two chordwise and three spanwise positions), 46 static pressure taps (34 on the suction side and 12 on the pressure side), and 36 unsteady pressure transducers (25 on the suction side in the central region, 5 distributed spanwise to capture 3D effects, and 6 on the pressure side).
The static pressure taps were located with a 10° outward or inward inclination to minimize cross-interference between neighboring taps.
Although this inclination was not strictly required for the present turbulent flow tests due to turbulence tripping, it was retained to accommodate potential future laminar flow investigations.
Four optical displacement sensors were positioned outside the test section to monitor rigid-body heave and pitch motions at the wing root attachment, and four additional accelerometers were used in selected test cases.

The experimental setup included a data acquisition chain connecting all unsteady sensors (accelerometers, unsteady pressure transducers, optical sensors), with data recorded using dedicated acquisition software.
Wind-tunnel measurements such as upstream static temperature and total pressure on both sides of the test section were acquired simultaneously.
All channels were synchronized and sampled at 1600 Hz (800 Hz bandwidth).

\begin{figure}
     \centering
         \includegraphics[width=\columnwidth]{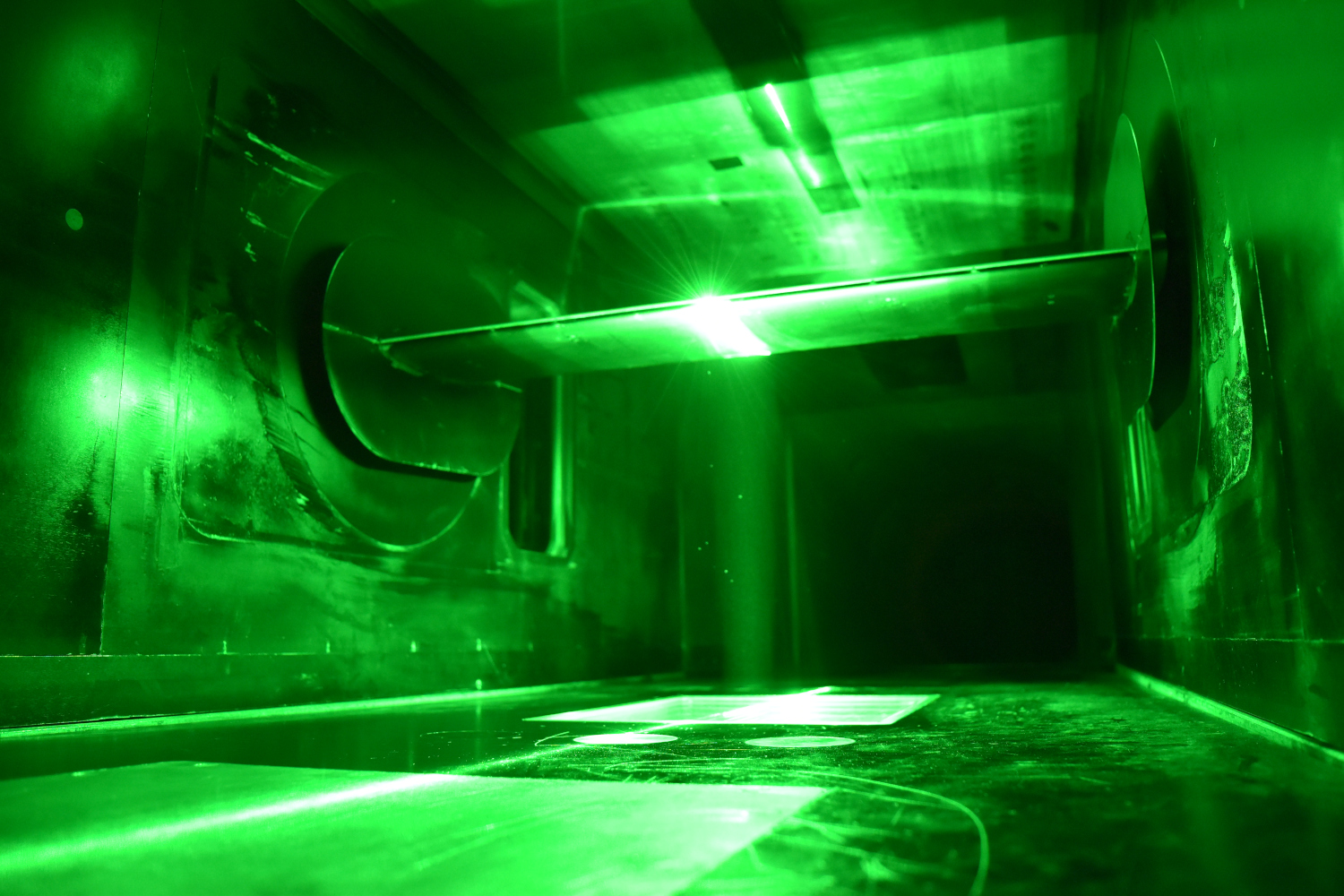}
        \caption{View of the OAT15A airfoil illuminated by the PIV laser sheet, inside S3Ch test section}
        \label{fig:model_in_S3Ch}
\end{figure}

Stereoscopic PIV was also implemented to obtain 2D-3C velocity measurements at mid-span on the suction side of the airfoil.
Since the laser sheet could only be introduced from the floor of the test section, the model was inverted so that the illuminated plane intersected the region of interest where shock buffet develops (see Fig.~\ref{fig:model_in_S3Ch}).
The adaptive upper wall then performed incidence and velocity distribution corrections on the inverted configuration by mirroring the aerodynamic ensemble about the lower wall.
While the PIV results are not analyzed in the present paper, the setup enabled time-resolved measurements of the flow dynamics in the shock and wake regions.
The acquisition and preliminary analysis of the PIV measurements from this campaign, together with the methodological challenges specific to moving-body configurations, are presented in~\cite{sicard2025experimental}.

\begin{figure}
    \centering
    \includegraphics[width=1\columnwidth]{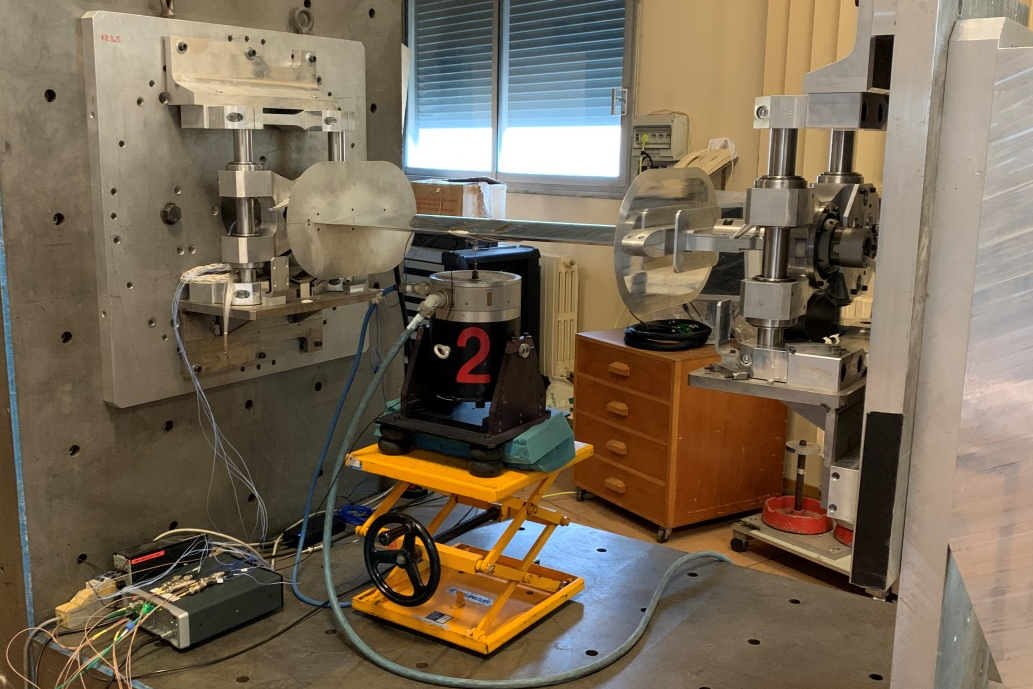}
    \caption{Experimental setup for preliminary wind-off tests in laboratory conditions}
    \label{fig:labtestsetup}
\end{figure}

\subsection{Structural dynamic identification of the model (wind-off tests)}\label{subsec:labtests}

\begin{figure*}[h]
    
        \centering
        \includegraphics[width=14cm]{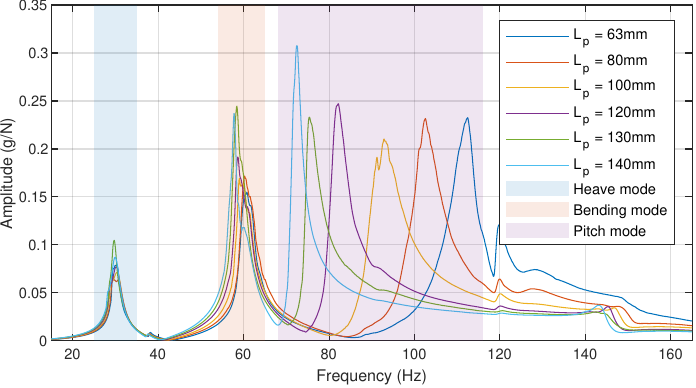}
    \caption{Evolution of structural frequencies and response amplitudes during wind-off tests, obtained from FRFs measured by a single accelerometer at $x/c = 0.9$, $y/b = 0.5$. The pitch stiffness setting $L_p$ is varied while the heave stiffness is held at setting 1.}
    \label{fig:labtestFRF}
\end{figure*}

\begin{figure*}[h]
    
        \centering
        \includegraphics[width=14cm]{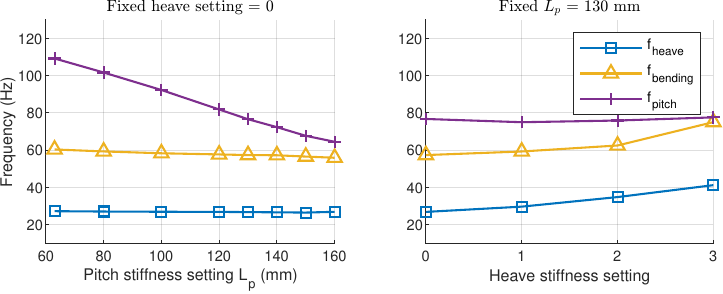}
   
    \caption{Evolution of structural frequencies during wind-off tests. Variation of pitch stiffness (left) and heave stiffness (right). Fixed parameters are indicated in each subplot.}
    \label{fig:labtest_pitch_plunge}
\end{figure*}

Preliminary wind-off tests were conducted to characterize the dynamic behavior of the model using the setup shown in Fig.~\ref{fig:labtestsetup}.
Fig.~\ref{fig:labtestFRF} presents the detailed Frequency Response Functions (FRFs), showcasing the effect of varying pitch stiffness at a fixed heave stiffness.
The blue, orange, and purple shaded regions indicate the frequency ranges associated with the heave, first out-of-plane bending, and pitch modes, respectively.
Extending this analysis to the heave DOF, Fig.~\ref{fig:labtest_pitch_plunge} summarizes the evolution of the natural frequencies of these three modes as stiffnesses are varied independently, spanning approximately 65--110~Hz and 25--42~Hz for the pitch and heave frequencies.
The results confirm a low level of coupling between the heave and pitch DOFs, as each can be tuned over a wide range with minimal influence on the other.
This decoupling fulfills a key design objective.

However, while the pitch frequency range met initial specifications, the achievable heave frequencies were lower than expected. This discrepancy highlights the structure's sensitivity to clamping boundary conditions, which deviate from the idealized assumptions in preliminary finite element models.
Similarly, the first bending mode ranged from 57 to 75~Hz depending on the heave setting, significantly below the target value.
This mode also exhibited a slight frequency decrease and increased response amplitude as the pitch and bending frequencies converged, suggesting a reduction in effective damping.
This behavior is consistent with literature regarding sensitivity to residual angular compliance about non-primary bearing axes.
Given the overlap between this bending mode frequency range and the predicted buffet frequency range, the test campaign required careful assessment of potential fluid-structure interactions.
Additionally, a rigid out-of-plane antisymmetric mode and an in-plane coplanar mode were observed with lower magnitudes at 38~Hz and 120~Hz, respectively, while the second out-of-plane bending mode at 150~Hz fell outside the range of interest.

\subsection{Aeroelastic stability of the test setup (flutter risk assessment)}\label{subsec:flutter_calculations}

\nomenclature{DLM}{Doublet-Lattice Method}
Preliminary aeroelastic analyses were carried out using Doublet-Lattice Method (DLM) flutter calculations (\cite{albano1969doublet,blair1992compilation}), using the modal dataset obtained during the laboratory campaign.
Although DLM provides only a linear and subsonic approximation of the aerodynamics, with known limitations in transonic conditions\footnote{The theoretical modeling is restricted to a linear representation of inviscid subsonic flow, and the behavior of the kernel function deteriorates at high subsonic Mach numbers as the flow approaches the transonic regime (\cite{richard1979program}).}, it remains suitable for a first-order stability screening.
The objective here was not to predict buffet-driven FSI, but simply to verify the absence of flutter within the operating envelope.

All investigated configurations exhibited positive modal damping throughout the tested range of Mach number and structural stiffnesses.
These results confirm that no flutter risk is expected within the range of conditions considered in the experimental matrix.
Still, given the inherent limitations of DLM in transonic flows, the wind-tunnel sequence was designed conservatively, beginning with the stiffest pitch configuration and gradually tuning down $f_{\text{pitch}}$ toward the predicted buffet frequency band.

\section{Wind tunnel test campaign}

\begin{table*}
\centering
\caption{Summary of test matrix}
\label{table:test_matrix}
\begin{tabular}{lccll}
\toprule
\textbf{Configuration} 	&	$\bm{\alpha_0}$ & \textbf{DOF settings} \\ 
\midrule
0 DOF &  0° to 4° & \makecell[l]{Heave and pitch blocked}  \\ \cmidrule{1-3}
1 DOF Heave &  2° to 4° &  \makecell[l]{Heave setting variations ($f_{\text{heave}}/f_{\text{buffet}} \approx $ 0.4 to 0.75)} \\  \cmidrule{1-3}
1 DOF Pitch	&  2° to 4° &  \makecell[l]{Pitch setting variations ($f_{\text{pitch}}/f_{\text{buffet}} \approx $ 1.5 to 2.2)}  \\  \cmidrule{1-3}
\makecell[l]{2 DOF} &  3.5° &  \makecell[l]{Pitch stiffness: $L_{p}$~=~100mm ($f_{\text{pitch}}/f_{\text{buffet}} \approx $ 1.7)\\ Heave stiffness: 3 (highest)  ($f_{\text{heave}}/f_{\text{buffet}} \approx $ 0.75) }	  \\ 
\botrule
\end{tabular}
\end{table*}

The S3Ch wind tunnel test campaign (see Fig.~\ref{fig:model_in_S3Ch}) covered several structural model configurations: 0 DOF (rigid setup), 1 DOF at a time (pitch or heave only), and 2 DOF simultaneously (pitch and heave). These will be referenced as 0 DOF, pitch (or 1 DOF pitch), heave (or 1 DOF heave), and 2 DOF configurations. For each configuration, the test matrix included variations in aerodynamic parameters (Mach number, AoA) and structural stiffness (heave/pitch DOF stiffness). Each parameter change required a new wind tunnel test run, except for Mach number. A summary of the test matrix is provided in Table~\ref{table:test_matrix}. Static taps, unsteady pressure sensors (Kulite), and accelerometer measurements were recorded. PIV measurements were performed for selected test cases.

The test campaign followed a sequential approach. First, the 0 DOF configuration was investigated to determine buffet occurrence across various Mach numbers and angles of attack (AoA), allowing comparison with previous experiments in the S3Ch wind tunnel (\cite{jacquin2009experimental}).
Subsequently, each degree of freedom was characterized independently to quantify its specific influence on buffet dynamics and fluid-structure interaction (FSI).
This exploratory phase revealed a lock-in phenomenon under specific structural and aerodynamic conditions.
Building upon these insights, a 2 DOF configuration exhibiting lock-in behavior was selected for in-depth analysis, with the objective of characterizing both transient and stabilized FSI phases.

Modal frequencies varied continuously with AoA and Mach number, similar to the shock buffet frequency, a behavior already noted in \cite{jacquin2009experimental}. To enable consistent comparison across aerodynamic conditions that inherently shift these frequencies, configurations were referenced by their prescribed stiffness setting rather than by the resulting natural frequency. For context, Table~\ref{table:test_matrix} reports the approximate DOF-to-buffet frequency ratios for the tested cases.

\begin{figure}
    \centering
    \includegraphics[width=\columnwidth]{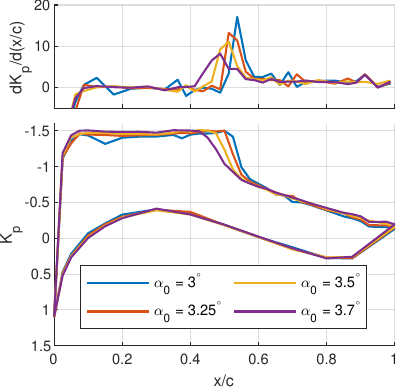}
    \caption{Mean surface pressure coefficient $K_p$ (bottom), and its chordwise derivative $dK_p/d(x/c)$ (top) along the suction side, for the 0 DOF configuration at $M = 0.70$ and $\alpha_0 = 3^\circ$--$3.7^\circ$.}
    \label{fig:Kp_Mach_0.70_with_derivative}
\end{figure}

\subsection{0-DOF configuration}
\label{subsec:0_DOF_conf}

In the 0 DOF configuration, the influence of angle of attack and Mach number on transonic shock buffet onset was investigated.
Fig.~\ref{fig:Kp_Mach_0.70_with_derivative} presents the mean surface pressure coefficient $K_p$ and its chordwise derivative for $M = 0.70$ and $\alpha_0 = 3^\circ$--$3.7^\circ$.

On the suction side, a pressure plateau characteristic of supercritical airfoils is observed up to mid-chord, followed by a sharp compression associated with the shock.
As $\alpha_0$ increases, the shock extends over a larger chordwise region, reflected in the progressive decrease of maximum $dK_p/d(x/c)$ values.
An upstream shift in the mean shock position and a spatially broadened shock region are visible for $\alpha_0 \geq 3.5^\circ$, suggesting buffet onset.
However, the gradual transition makes it difficult to identify a precise buffet onset angle from mean surface pressure coefficients alone.

\begin{figure}
    \centering
    \includegraphics[width=\columnwidth]{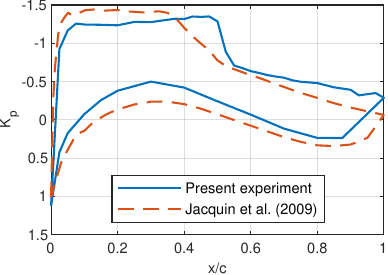}
    \caption{Comparison of mean surface pressure coefficient $K_p$ between the present experiment and \cite{jacquin2009experimental} for $M_0 = 0.73$ and $\alpha_0 = 3.5^\circ$.}
    \label{fig:comparatif_HOMER_Jacquin}
\end{figure}

\begin{figure*}[h]
    \centering
    \includegraphics[width=\textwidth]{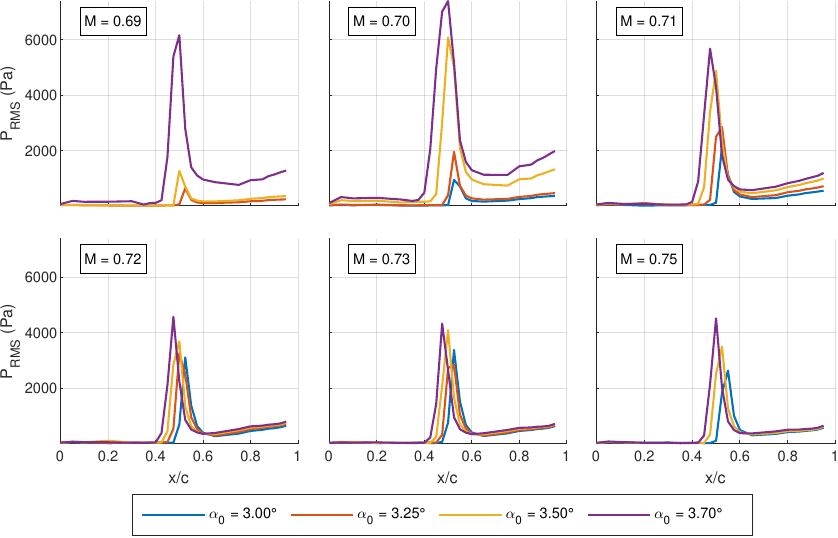}
    \caption{RMS of unsteady pressure sensors versus $x/c$ for various Mach numbers and $\alpha_0$. Note: not all $\alpha_0$ values are present for every Mach number.}
    \label{fig:RMS_Kulite_par_Mach}
\end{figure*}

\begin{figure}
    \centering
        \includegraphics[width=\columnwidth]{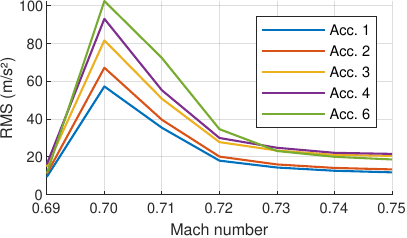}
        \caption{RMS of accelerometer measurements versus Mach number at $\alpha_0 = 3.5^\circ$. Accelerometer 5 was damaged and excluded from the analysis.}
    \label{fig:Run5899_RMS_accelero}
\end{figure}

Given the similarities between the present test campaign and those conducted by \cite{jacquin2009experimental,jacquin2016testing}, notably the use of the same wind tunnel and OAT15A wing profile, a direct comparison was performed for $M_0 = 0.73$ and $\alpha_0 = 3.5^\circ$ (Fig.~\ref{fig:comparatif_HOMER_Jacquin}).
In Jacquin's data, the shock/recompression region extends significantly along the chord ($x/c \approx 0.35$--$0.5$), suggesting temporal shock oscillations.
Such buffet characteristics are not observed in the present experiment at this Mach number.
Buffet onset occurs at lower Mach numbers and vanishes at higher Mach numbers where Jacquin et al. still reported its presence.

Several main factors are therefore believed to account for the discrepancies.
First, the wall interference treatment differs: Jacquin's campaigns used adaptive walls on both sides, whereas the present experiment used only the upper adaptive wall.
Second, structural flexibility plays a role: even in its nominally rigid configuration, the present setup allows two degrees of freedom for FSI studies.
The no-contact, labyrinth-based seal at the wall interface leads to clamping the wing further from mid-span, providing greater bending flexibility.
As a result, slight FSI effects may have occurred, which are known to lower the buffet threshold (\cite{gao2020transonic}).
Variations in the effective angle of attack under aerodynamic loading were investigated but ruled out: optical sensor measurements indicated only small angular deflections ($0.01^\circ$--$0.03^\circ$).

\begin{figure*}
    \centering
    \begin{subfigure}[b]{\textwidth}
        \includegraphics[width=\textwidth]{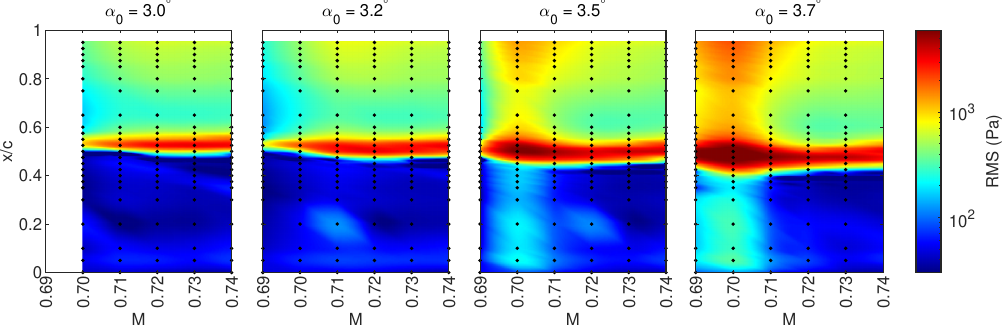}
        \caption{Root Mean Square (RMS) of unsteady pressure.}
        \label{fig:RMS_Kulite_alpha_surface}
    \end{subfigure}
    \vspace{0.5cm}
    \begin{subfigure}[b]{\textwidth}
        \includegraphics[width=\textwidth]{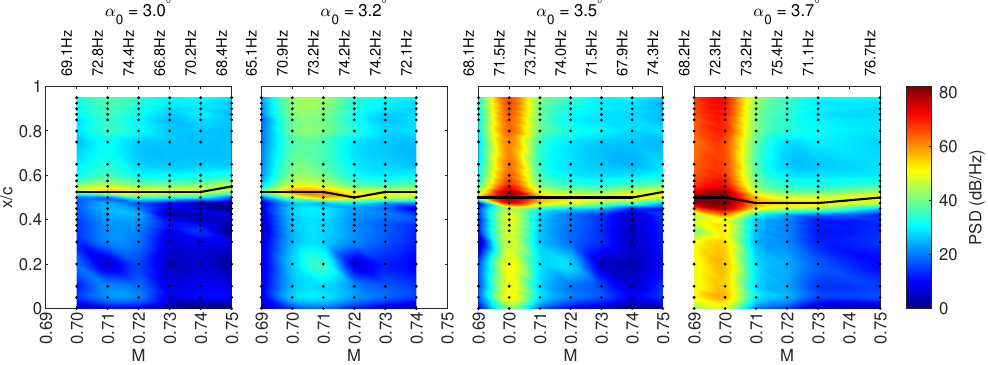}
        \caption{Power Spectral Density (PSD) of unsteady pressure at the frequency of maximum amplitude. Vertical labels indicate the spectral peak frequency for each Mach number. Black lines track the shock location (sensor with maximum amplitude).}
        \label{fig:PSD_Kulite_alpha}
    \end{subfigure}
    \caption{RMS (top) and PSD (bottom) values of unsteady pressure sensors along the airfoil chord ($x/c$) versus Mach number ($M$) for four angles of attack ($\alpha_0$). Dots: unsteady pressure sensor locations. Color surface: cubic interpolation. Each figure shares a common logarithmic color scale.}
    \label{fig:RMS_PSD_Kulite_alpha}
\end{figure*}

\begin{figure}
    \centering
        \includegraphics[width=\columnwidth]{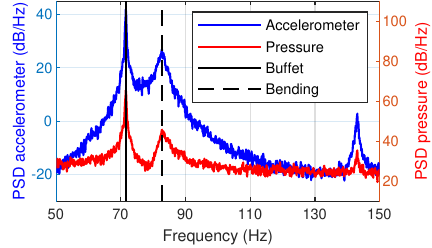}
        \caption{PSD of accelerometer ($x/c=0.1$, $y/b=0.5$) and unsteady pressure sensor ($x/c=0.925$, $y/b=0.43$) at $M_0~=~0.70$ and $\alpha_0 = 3.5^\circ$.}
    \label{fig:Run5899_PSD_accelero3_kulite24}
\end{figure}

Building on the previous analysis of the static pressure distributions, including shock locations and their chordwise extent, the unsteady surface pressures were then analyzed to characterize the flow dynamics.
To this end, Root Mean Square (RMS) levels and Power Spectral Densities (PSD) from the unsteady pressure sensors were examined to detect unsteady phenomena and potential buffet signatures.
\nomenclature{RMS}{Root Mean Square}
\nomenclature{PSD}{Power Spectral Density}

The evolution of RMS values across unsteady pressure sensors and accelerometers quantifies the global aerodynamic and structural unsteadiness.
Fig.~\ref{fig:RMS_Kulite_par_Mach} presents the RMS of pressure fluctuations on the suction side.
Two main trends emerge.
At $M=0.69$, fluctuations remain weak for $\alpha_0=3.25^\circ$ and $3.5^\circ$ but increase sharply at $\alpha_0=3.7^\circ$, particularly near the shock and trailing-edge regions, indicating buffet onset.
A similar behavior is observed at $M=0.70$, although buffet occurs earlier ($\alpha_0=3.5^\circ$) and with greater intensity.
In contrast, for $M\ge 0.72$, fluctuation amplitudes remain low and nearly independent of Mach number and AoA.
The case $M=0.71$ represents a transitional regime where buffet, already present at lower Mach numbers, begins to subside.
A moderate but progressive decrease in unsteady pressure levels is observed for $\alpha_0=3.5^\circ$ and $3.7^\circ$, suggesting buffet attenuation rather than a fully developed buffet.

A consistent trend is observed in the structural response (Fig.~\ref{fig:Run5899_RMS_accelero}).
RMS values measured by the accelerometers at $\alpha_0=3.5^\circ$ remain weak at $M=0.69$ and for $M\ge 0.72$, but rise markedly at $M=0.70$ and, to a lesser extent, at $M=0.71$.
This behavior aligns with the onset and subsequent attenuation of buffet-induced oscillations observed in the pressure data.

These findings are summarized in Fig.~\ref{fig:RMS_Kulite_alpha_surface}, which shows the measured and interpolated RMS pressure fluctuations as a function of Mach number (horizontal axis) and chordwise position (vertical axis) for four $\alpha_0$ values.
Buffet onset at $\alpha_0 = 3.5^{\circ}$ and $3.7^{\circ}$ is evident from the pronounced increase in unsteady fluctuations with AoA, particularly in the shock region and trailing-edge area.

While RMS values provide a global overview of unsteady behavior across the entire frequency range, PSD analysis allows specific frequency components, such as the buffet frequency, to be isolated and examined.
For each test case, the PSD amplitude at the frequency of maximum spectral peak in the buffet frequency range was extracted for all pressure sensors.

Similarly to the RMS map, Fig.~\ref{fig:PSD_Kulite_alpha} presents the PSD amplitudes of unsteady pressure sensors along the airfoil chord for several Mach numbers and angles of attack.
A pronounced maximum is observed at $M=0.70$, for $\alpha_0 = 3.5^\circ$ and $3.7^{\circ}$, with a spatial distribution consistent with the RMS patterns shown in Fig.~\ref{fig:RMS_Kulite_alpha_surface}.
The coherence between the broadband and spectral responses indicates that the fluctuations at the buffet frequency dominate the overall unsteady pressure field.
The subsequent reduction in amplitude at higher Mach numbers reflects a transition toward a more stable aerodynamic regime with limited shock motion.

To further characterize the FSI, Fig.~\ref{fig:Run5899_PSD_accelero3_kulite24} compares the PSD of one selected accelerometer and unsteady pressure sensor at $M_0 = 0.70$ and $\alpha_0 = 3.5^\circ$.
The spectral peaks corresponding to the buffet and bending frequencies (at 71~Hz and 83~Hz respectively), along with the first harmonic of the buffet, coincide for both measurements.
This agreement confirms that the structure responds directly to the aerodynamic buffet excitation, while the pressure sensor also captures the structural bending mode.

\begin{figure}
    \centering
    \includegraphics[width=\columnwidth]{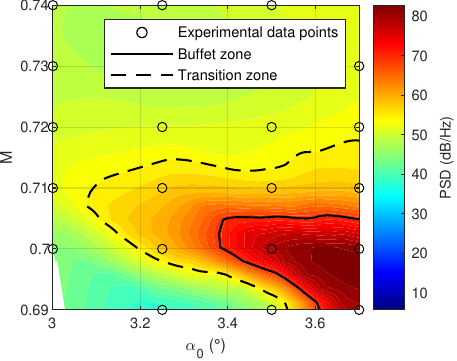}
    \caption{Experimental and interpolated maximum amplitudes of PSD at buffet frequency on unsteady pressure sensors, and proposed buffet and transition zones,
     depending on $\alpha_0$ and Mach number.}
    \label{fig:MaxAmplitude_3D_Scatter}
\end{figure}

\begin{figure}
    \centering
    \includegraphics[width=\columnwidth]{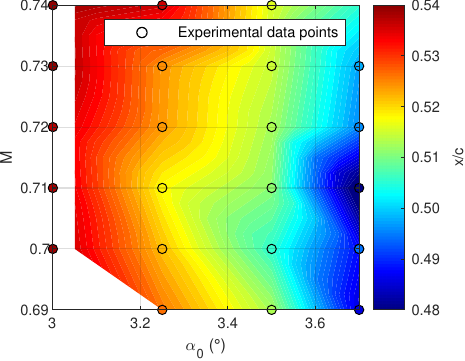}
    \caption{Experimental and interpolated location in chord (x/c) of the shock position, based on static pressure taps pressure distribution gradients, depending on $\alpha_0$ and Mach number.}
    \label{fig:SensorLocation_3D_Scatter}
\end{figure}

A synthesis of these experimental findings is presented in Fig.~\ref{fig:MaxAmplitude_3D_Scatter}, which illustrates the variation of dynamic response amplitude as a function of Mach number and angle of attack.
The results reveal a broad transitional region where the flow exhibits intermediate pre-buffet behavior prior to the full establishment of buffet.
The delineation of the buffet zone should therefore not be interpreted as a sharp boundary, but rather as an indicative region where buffet oscillations are clearly established.
Complementary analyses conducted in~\cite{sicard2025experimental}, based on an energy-based approach, led to conclusions which are consistent with the present results.

\cite{accorinti2022experimental} also proposed a methodology for identifying buffet onset based on shock-position tracking, with particular emphasis on detecting shock-motion inversion, a phenomenon recognized as a precursor to buffet onset (\cite{pearcey1955some}) and typically followed by amplified unsteady flow fluctuations (\cite{lee2001self,jacquin2009experimental}).
Their approach employed Background-Oriented Schlieren (BOS) imaging.
In this experiment, the mean shock position was inferred from the static-pressure tap line located near mid-span.
From the suction-side pressure-coefficient distribution $Kp(x/c)$ (see Fig.~\ref{fig:Kp_Mach_0.70_with_derivative}), the streamwise gradient $dK_p / d(x/c)$ was evaluated between adjacent taps.
A linear interpolation of this gradient along the chord was then performed to locate its extremum, which corresponds to the chordwise coordinate $x/c$ of the shock.
The results are summarized in Fig.~\ref{fig:SensorLocation_3D_Scatter}, which maps the inferred shock locations in the Mach-AoA plane.
A clear trend emerges: the shock moves upstream as the AoA increases and buffet sets in.
This "inverse shock motion" agrees with the numerical results of \cite{nitzsche2019fluid} on the OAT15A airfoil in a rigid configuration.
At higher Mach numbers ($M \geq 0.72$) though, for $\alpha_0~=~3.5^\circ$ and $3.7^\circ$, the shock appears to shift downstream again, an indicator of offset, in accordance with the buffet occurrence map observed in Fig.~\ref{fig:MaxAmplitude_3D_Scatter}.

To summarize, the flow exhibits either fully developed buffet or transitional unsteady behavior for Mach numbers between approximately $M~=~0.69$ and $0.71$, and AoA starting from $\alpha_0~\approx~3.1/3.2^\circ$ (Fig.~\ref{fig:MaxAmplitude_3D_Scatter}).
Within this range, regions of fully developed buffet coexist with areas where the flow displays only intermittent or evolving unsteadiness, reflecting a gradual onset of the phenomenon rather than a sharp threshold.
The dominant buffet frequency lies between approximately 68~Hz and 75~Hz (Fig.~\ref{fig:RMS_PSD_Kulite_alpha}), consistent with the findings of \cite{jacquin2009experimental} when accounting for the difference in airfoil chord length.
The amplitude of the unsteady response generally increases with AoA, indicating progressive intensification as the flow transitions toward fully established buffet.
Buffet consistently ceases at higher Mach numbers for $M\geq~0.72$ approximately.

\FloatBarrier

\subsection{DOF-enabled configurations}
\label{subsec:dof_enabled}

\subsubsection{1 DOF pitch configuration}
\label{subsubsec:pitch_conf}

The 1 DOF pitch testing campaign explored the effects of AoA and Mach number on buffet onset and FSI for various pitch stiffness settings.
AoA ranging from $2^\circ$ to $4^\circ$ were tested, with a focus on the buffet onset region ($\alpha_0 = 3^\circ$ to $4^\circ$).
Mach number was varied from 0.50 to 0.75, with particular attention to the range $M = 0.68$--$0.72$ where buffet activity was expected.
Several pitch stiffness configurations were tested by varying the pitch stiffness length $L_p$ from 63~mm (stiffest) to 100~mm (most flexible), corresponding to a pitch-to-buffet frequency ratio $f_{\text{pitch}}/f_{\text{buffet}}$ decreasing from approximately 2.2 ($L_p = 63$~mm) to 1.5 ($L_p = 100$~mm).
Stiffness was progressively reduced in chronological order to approach resonance between the pitch mode and buffet frequencies.

Specific testing regimes revealed pronounced FSI, characterized by the emergence of LCO and a lock-in phenomenon.
The main trends indicate that the amplitude of FSI intensified with increasing angle of attack (AoA) and as the pitch frequency decreased (achieved by increasing $L_{\text{pitch}}$), approaching the buffet frequency.

\begin{figure}
     \centering
     \begin{subfigure}[b]{\columnwidth}
         \centering
                 \includegraphics[width=\textwidth]{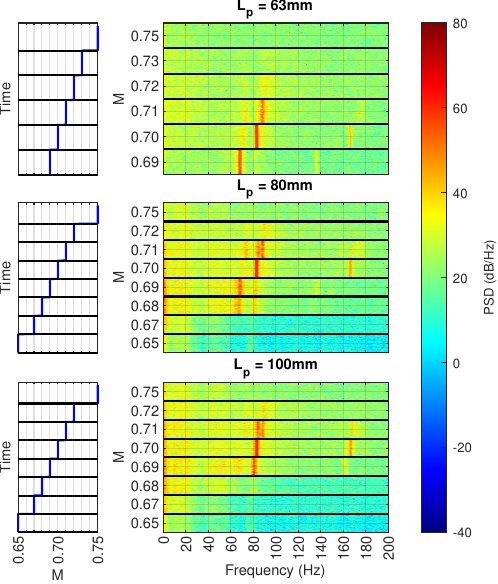}
         \captionsetup{justification=centering}
         \caption{Unsteady pressure sensor ($x/c=0.925$, $y/b=0.43$)}
        \label{fig:1DOF_pitch_spectrogram_KE24}
     \end{subfigure}
     \begin{subfigure}[b]{\columnwidth}
         \centering
        \includegraphics[width=\textwidth]{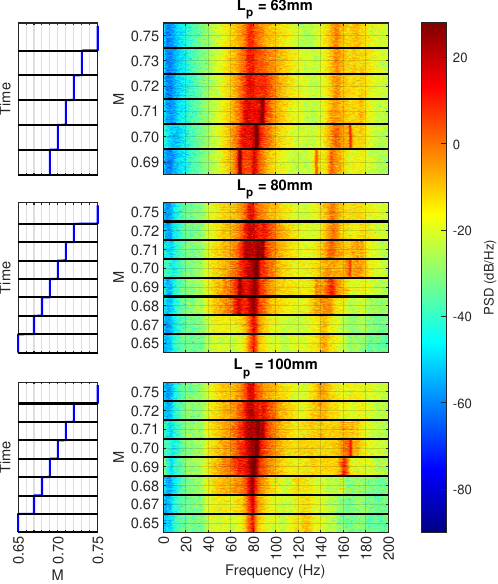}
        \caption{Accelerometer ($x/c=0.9$, $y/b=0.5$)}
        \label{fig:1DOF_pitch_spectrogram_acc4}
     \end{subfigure}
    \caption{Spectrograms of unsteady pressure and accelerometer signals at successive stabilized Mach numbers for three pitch stiffness settings ($L_{\text{pitch}} = 63$, 80, and 100~mm). Rows correspond to stiffness values from stiffest (top) to most flexible (bottom). The y-axis shows the Mach number for each stabilized test condition (time axis relabeled). 1 DOF pitch configuration, $\alpha_0=3.5^\circ$.}
        \label{fig:1DOF_pitch_spectrogram}
\end{figure}

Fig.~\ref{fig:1DOF_pitch_spectrogram} presents spectrograms of unsteady pressure (Kulite) and accelerometer signals, both located near the mid-span trailing edge. Data were acquired across several Mach numbers for three pitch stiffness settings, ranging from the stiffest ($L_p = 63$~mm) to the most flexible ($L_p = 100$~mm), at a constant angle of attack $\alpha_0 = 3.5^\circ$.
Signals were sampled at $F_s = 1600$~Hz, with each stabilized Mach increment recorded over approximately 1~minute to ensure steady-state conditions.
The Power Spectral Density (PSD) was estimated using the Short-Time Fourier Transform (STFT) with a Hann window of length $N=1600$ samples ($\Delta t = 1$~s), yielding a frequency resolution of $\Delta f = 1$~Hz.
An 80\% overlap between segments was applied to improve temporal smoothness.
PSD values were normalized by the window energy and sampling frequency to obtain physical units of $\text{Pa}^2/\text{Hz}$ for pressure and $(m/s^2)^2/\text{Hz}$ for acceleration, then converted to a logarithmic scale (dB/Hz).
To enhance visual continuity across the test matrix, successive stabilized Mach segments were concatenated for analysis.
Although the underlying data is time-based, the vertical axis of the spectrograms represents the corresponding Mach number for each test condition rather than absolute time.
Note that time intervals corresponding to flow stabilization and wind tunnel adaptation between Mach increments are excluded from the visualization.

The unsteady-pressure sensor spectrograms (Fig.~\ref{fig:1DOF_pitch_spectrogram_KE24}) reveal a buffet signature in the 65--70~Hz band, consistent with the predicted buffet-frequency range, for $L_p = 63$~mm.
The same feature is visible for $L_p = 80$~mm at $M=0.69$, and at $M=0.68$ with weaker magnitude.\footnote{Mach numbers below $M=0.69$ could not be tested for $L_p = 63$~mm.}
Increasing the Mach number to $M=0.70$ yields a frequency shift above 80~Hz and increased unsteady response magnitude, characteristic of lock-in, for both settings.
Notably, $L_p = 63$~mm exhibits a clear shift between $M=0.69$ and $M=0.70$, while $L_p = 80$~mm displays transitional behavior at $M=0.68-0.69$, with peaks at both buffet (weaker) and, albeit weaker, at the lock-in (or bending mode, as discussed below) frequencies.
In contrast, $L_p = 100$~mm shows direct onset of buffet at the lock-in frequency, even at $M=0.69$.
Therefore, it appears that decreasing the pitch stiffness for identical aerodynamic parameters ($M$, $\alpha_0$) decreases the response amplitude at the buffet frequency while increasing it at the lock-in frequency.

\begin{figure*}[htbp]
    \centering
    \includegraphics[width=\textwidth]{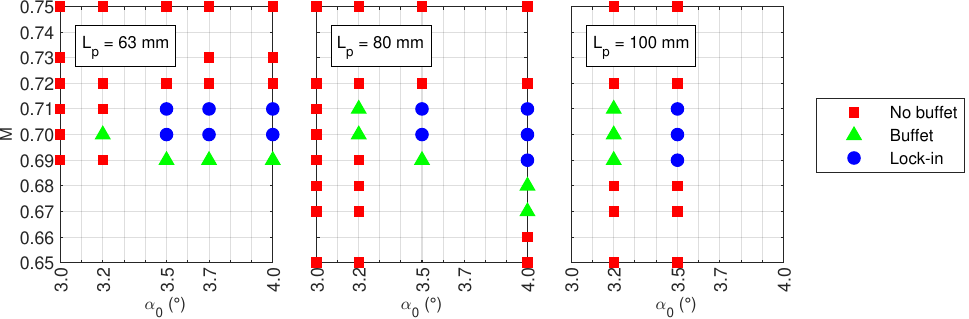}
    \caption{Buffet and lock-in occurrence maps as a function of Mach number and angle of attack for three pitch stiffness settings (1 DOF pitch configuration).}
    \label{fig:Buffet_maps_1DOFpitch}
\end{figure*}

It can also be noted that small Mach number increments (from $M=0.69$ to $M=0.71$) slightly increase the lock-in frequency, which disappears at $M=0.72$ and above for all stiffness settings.
Harmonic content at twice the fundamental frequencies is also visible in all figures.

The structural response, visible in Fig.~\ref{fig:1DOF_pitch_spectrogram_acc4}, provides complementary information, while being consistent with the aerodynamic observations.
A clear peak near 80~Hz is observable in all three spectrograms, which corresponds to the first bending mode, identified through modal analysis.
Since the lock-in frequency lies essentially on top of this bending frequency, the data suggest a coupling between the buffet dynamics and the structural bending mode.
The lock-in frequency does increase when increasing Mach number though, while the bending mode frequency appears to remain constant.
Also identified via modal analysis, the pitch mode is observable in the spectrograms at approximately 150~Hz, 140~Hz and 130~Hz for the stiffness settings $L_p = 63$, $80$ and $100\,\text{mm}$, respectively.
For the stiffer configurations ($L_p = 63$--80~mm), this natural pitch frequency approaches the region of the first harmonic of the buffet-bending lock-in (around 160~Hz).
Consequently, when lock-in is active the pitch mode peak is sometimes overshadowed by the harmonic component of the bending or lock-in frequencies.

A series of AoA/Mach-number tests was carried out for the three pitch-stiffness settings with the aim of characterizing buffet and possible lock-in.
Concatenated spectrograms (as seen in Fig.~\ref{fig:1DOF_pitch_spectrogram} for $\alpha_0=3.5^\circ$) were examined for all cases and AoA values.
From these analyses, buffet and lock-in occurrence maps were generated for the three pitch-stiffness configurations and are shown in Fig.~\ref{fig:Buffet_maps_1DOFpitch}.

In most configurations, as in the 0-DOF reference case (Fig.~\ref{fig:MaxAmplitude_3D_Scatter}), buffet occurs for Mach numbers $M = 0.69$--$0.71$, and as low as $M = 0.67$ for $\alpha_0 = 4^\circ$ and $L_p = 80$~mm.
Lock-in is consistently detected at $M = 0.70$ and $0.71$, and also appears at $M = 0.69$ for $\alpha_0 = 4^\circ$ and $L_p = 80$~mm.
The unsteady-pressure spectra indicate that the dynamic response consistently reaches its maximum at $M = 0.70$, as observed in Fig.~\ref{fig:1DOF_pitch_spectrogram}.

During the campaign, the large dynamic response of the model, especially under lock-in, caused the failure of several Kulite sensors (notably sensors located at the nominal shock position).
Because the response amplitude increases with both $\alpha_0$ and decreasing $f_{\text{pitch}}/f_{\text{buffet}}$ ratios, fatigue damage eventually constrained the test campaign.
Following a screw failure in the pitch-setting assembly, further tests were therefore limited to $\alpha_0 = 3.5^\circ$, a condition where buffet and lock-in were already well established.

The amplification of the response as the pitch and buffet frequencies converge is consistent with the numerical predictions of \cite{giannelis2016dynamic}.
The lowest tested ratio ($f_{\text{pitch}}/f_{\text{buffet}} \approx 1.5$) corresponds to the regime where FSI is expected to be strongest.
According to~\cite{giannelis2016dynamic}, closer alignment of $f_{\text{pitch}}$ with $f_{\text{buffet}}$ could theoretically lead to response attenuation beyond this point.
However, the numerical study considered a two-dimensional wing and did not include the three-dimensional out-of-plane bending mode, whose natural frequency lies between the buffet and pitch frequencies and likely plays a central role in the observed lock-in behavior.
Consequently, existing simulations cannot determine whether further frequency convergence would mitigate or intensify FSI.
A simultaneous alignment of buffet, bending, and pitch frequencies could instead amplify the response and threaten the structural integrity of the rig, which is why such configurations were not pursued in this campaign.

Ultimately, the addition of the pitch DOF changes the dynamic response of the system drastically: the transonic buffet and the bending mode, already present in the 0~DOF configuration, interact with the newly introduced pitch mode, giving rise to a fluid-structure lock-in phenomenon whose dominant frequency coincides with the bending frequency.

\subsubsection{1 DOF heave configuration}
\label{subsubsec:heave_conf}

\begin{figure}
     \centering
     \begin{subfigure}[b]{\columnwidth}
         \centering
         \includegraphics[width=\textwidth]{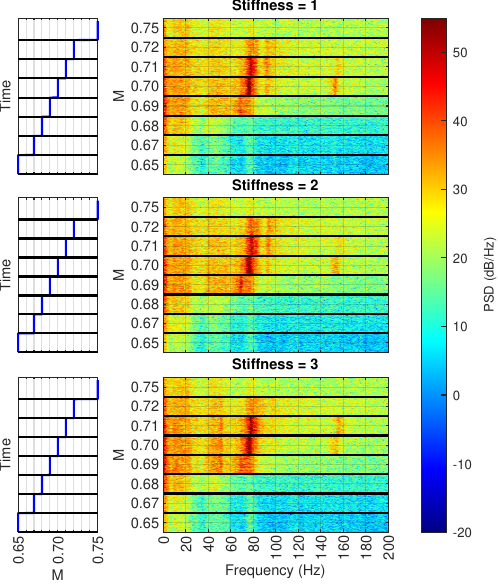}         
         \captionsetup{justification=centering}
        \caption{Unsteady pressure sensor ($x/c=0.925$, $y/b=0.43$)}
        \label{fig:1DOF_heave_spectrogram_KE24}
     \end{subfigure}     
     \begin{subfigure}[b]{\columnwidth}
         \centering
         \includegraphics[width=\textwidth]{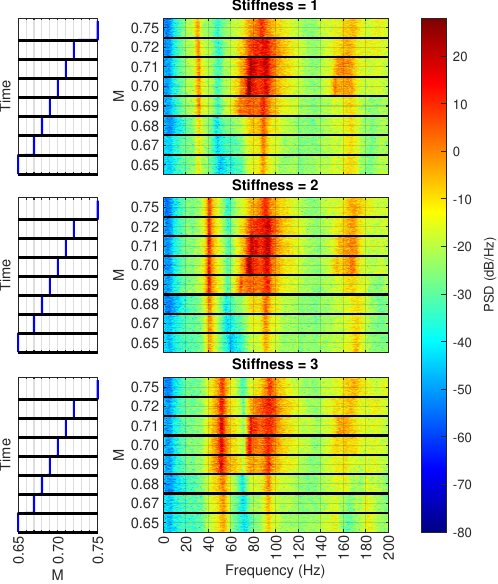}       
        \caption{Accelerometer ($x/c=0.9$, $y/b=0.5$)}  
        \label{fig:1DOF_heave_spectrogram_acc4}
     \end{subfigure}
     \caption{Spectrograms of unsteady pressure and accelerometer signals at successive stabilized Mach numbers for three heave stiffness settings (1, 2, and 3), from most flexible (top) to stiffest (bottom). The y-axis shows the Mach number for each stabilized test condition (time axis relabeled). 1~DOF heave configuration, $\alpha_0 = 3.5^\circ$.}
        \label{fig:1DOF_heave_spectrogram}
\end{figure}

Experimental tests were performed on the 1 DOF heave configuration while varying angle of attack ($\alpha_0 = 3^\circ$, $3.2^\circ$, $3.5^\circ$), Mach number, and heave stiffness settings (low, medium, high; see Fig.~\ref{fig:DOF_test_setup}).
As in the rigid 0 DOF case, the pitch mode remained present but its frequency was identified via modal analysis at approximately 170~Hz, giving $f_{\text{pitch}}/f_{\text{buffet}} \geq 2.3$.
This ratio is safely above the maximum lock-in range ratio of 1.8 reported in~\cite{giannelis2016dynamic}, and thus unlikely to affect the observed fluid-structure interaction.

Spectrograms for the same sensor pair and the three stiffness levels are displayed in Fig.~\ref{fig:1DOF_heave_spectrogram}.
The heave frequency rises with stiffness (32~Hz to 52~Hz), while the buffet frequency remains essentially constant at $f_{\text{buffet}} \approx 77$~Hz for all cases.
A modest increase of the bending frequency (from $\approx 90$~Hz to 95~Hz) is also observed, likely caused by the altered aerodynamic loading on the bearing supports.

No lock-in between buffet and any structural mode is detected; the buffet peak does not migrate toward the bending mode or the heave mode, even when the heave frequency is set to its highest setting.
Furthermore, a comparative analysis of 0 DOF and 1 DOF heave configurations across other Mach numbers and AoA revealed a subtle delay in buffet onset for the 1 DOF heave configuration, occurring at $\alpha_0 = 3.5^\circ$ instead of $\alpha_0 = 3.25^\circ$ at $M = 0.70$.

The key observation for the 1 DOF heave configuration is that the buffet frequency remains essentially unchanged and never locks in to the bending frequency at any Mach number or stiffness setting.
By contrast, the 1 DOF pitch configuration (Fig.~\ref{fig:1DOF_pitch_spectrogram}) did exhibit lock-in.
These results indicate that the mere proximity of the bending mode to the buffet frequency is insufficient to trigger lock-in.
A pitch DOF with a natural frequency sufficiently close to the buffet frequency appears to be a necessary condition.

\subsubsection{2 DOF pitch and heave configuration}
\label{subsubsec:2_DOF_conf}

Building on the single-DOF results, a coupled pitch-heave configuration was selected with $\alpha_{0}=3.5^{\circ}$. The pitch DOF stiffness was set to $L_{p}=100\,\text{mm}$ (corresponding to the lowest pitch natural frequency investigated in §\ref{subsubsec:pitch_conf}), while the heave DOF stiffness was set to 3 (the highest available).
By bringing both DOF natural frequencies as close as possible to the buffet range, within safe operational limits, this configuration ensures frequency proximity, making it well suited for examining the onset, development, and decay of fluid-structure coupling and frequency lock-in.

Extended continuous runs were performed over the narrow Mach-number interval encompassing buffet onset, lock-in, and buffet cessation.
The Mach number was first set below buffet onset, then increased to trigger buffet and FSI coupling, and finally varied both upward and downward to suppress the buffet and return to a steady state.
Complementary steady-state measurements were also performed at discrete Mach points, following the procedure used in the single-DOF campaigns.

\begin{figure}
     \centering
    \begin{subfigure}[b]{\columnwidth}
         \centering
         \includegraphics[width=\textwidth]{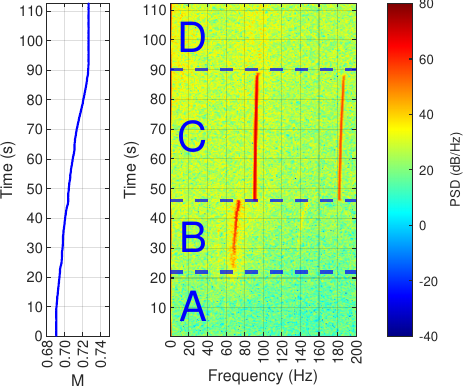}
         \captionsetup{justification=centering}
        \caption{Unsteady pressure sensor ($x/c=0.925$, $y/b=0.43$)}
        \label{fig:Spectrogram_kulite_comments}
     \end{subfigure}
     \hfill
     \begin{subfigure}[b]{\columnwidth}
         \centering
         \includegraphics[width=\textwidth]{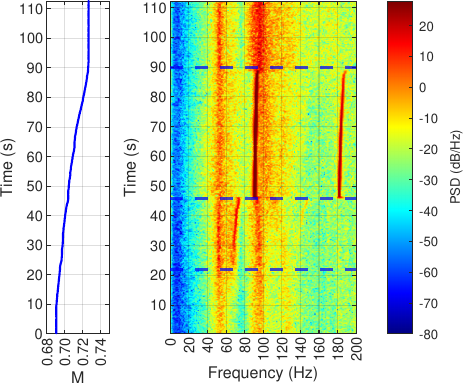}
        \caption{Accelerometer ($x/c=0.1$, $y/b=0.5$)}         
        \label{fig:Spectrogram_accelero}
     \end{subfigure}
     \hfill

        \caption{Spectrograms showing stages A (pre-buffet), B (buffet development), C (lock-in) and D (post lock-in) of transient behavior for $\alpha_0 = 3.5^\circ$  and varying Mach number.}
        \label{fig:Spectrograms}
\end{figure}

Fig.~\ref{fig:Spectrograms} shows the PSD spectrograms of a trailing-edge pressure sensor and a mid-span leading-edge accelerometer.
Time is represented on the vertical axis, corresponding to the actual elapsed time of a single continuous run (as opposed to the concatenated steady-state points used for the 1-DOF cases).
The left-hand subplot displays the Mach number evolution, while the main panels capture the full transient evolution of the aero-elastic system.

It is worth noting that, during these transient runs, the Mach number could not be continuously corrected for wall-interference and blockage effects.
Consequently, the Mach-number values obtained in the transient runs may differ slightly from those of the steady-state cases and should be interpreted qualitatively, with emphasis on the observed flow phenomenology rather than on precise quantitative comparison.

\begin{figure*}[ht]
    \centering
    \captionsetup[subfigure]{labelformat=empty} 

    \textbf{(A) Before buffet onset}
    \label{fig:before_buffet}\\[0.3em]
    \begin{subfigure}{0.32\textwidth}
        \includegraphics[width=\linewidth]{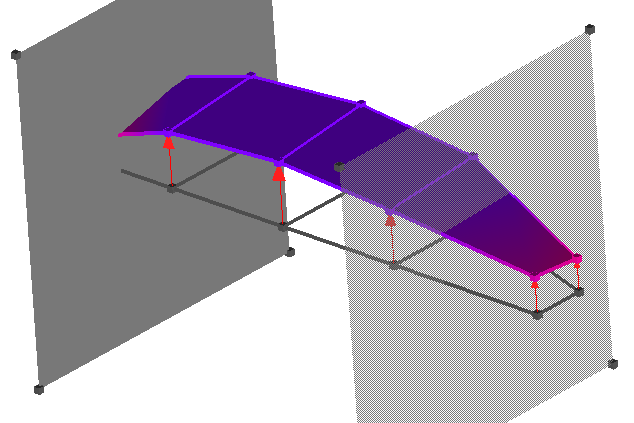}
        \caption{Heave mode (52~Hz)}
    \end{subfigure}
    \hfill
    \begin{subfigure}{0.32\textwidth}
        \includegraphics[width=\linewidth]{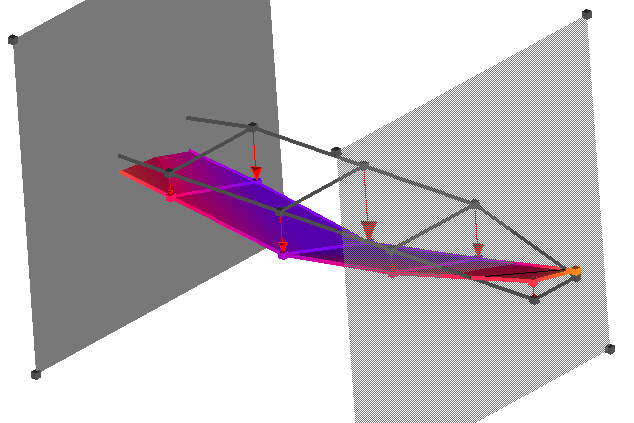}        
        \caption{Bending mode (96~Hz)}
    \end{subfigure}
    \hfill
    \begin{subfigure}{0.32\textwidth}
        \includegraphics[width=\linewidth]{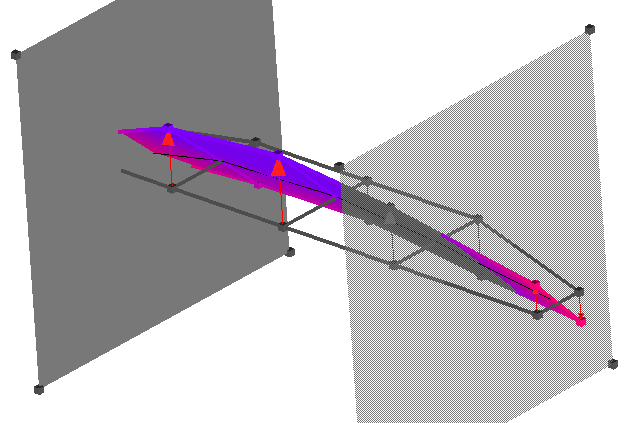}
        \caption{Pitch mode (125~Hz)}
    \end{subfigure}

    \vspace{1em}

    \textbf{(B) With buffet}
    \label{fig:with_buffet}\\[0.3em]
    \begin{subfigure}{0.32\textwidth}
        \includegraphics[width=\linewidth]{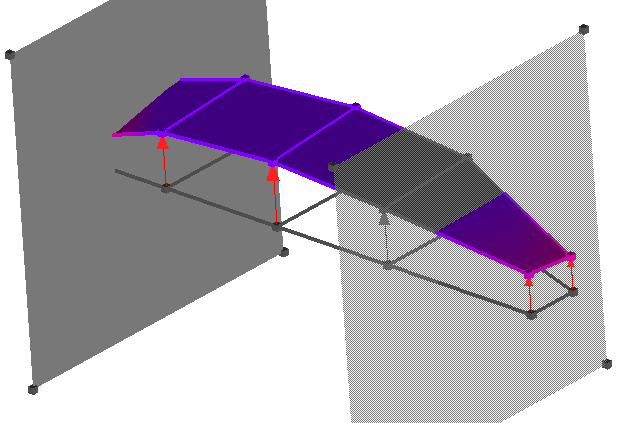}
        \caption{Heave mode (52~Hz)}
    \end{subfigure}
    \hfill
    \begin{subfigure}{0.32\textwidth}
        \includegraphics[width=\linewidth]{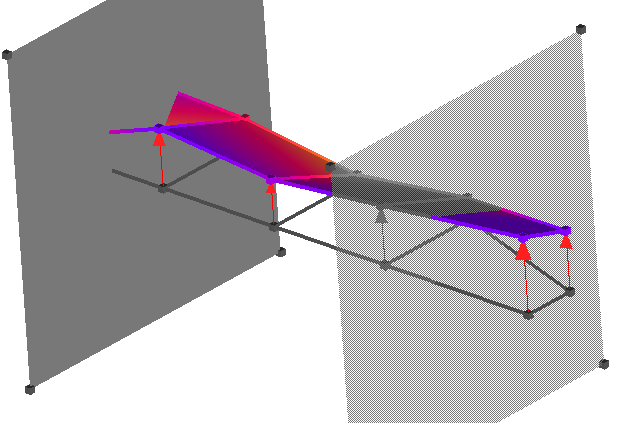}
        \caption{Buffet mode (70~Hz)}
    \end{subfigure}
    \hfill
    \begin{subfigure}{0.32\textwidth}
        \includegraphics[width=\linewidth]{Figures/5994-2-Buffet-Mode1-Heave-52Hz-greywall-cropped.png}
        \caption{Bending mode (95~Hz)}
    \end{subfigure}

    \vspace{1em}

    \textbf{(C) With lock-in}
    \label{fig:with_lockin}\\[0.3em]
    \begin{subfigure}{0.32\textwidth}
        \includegraphics[width=\linewidth]{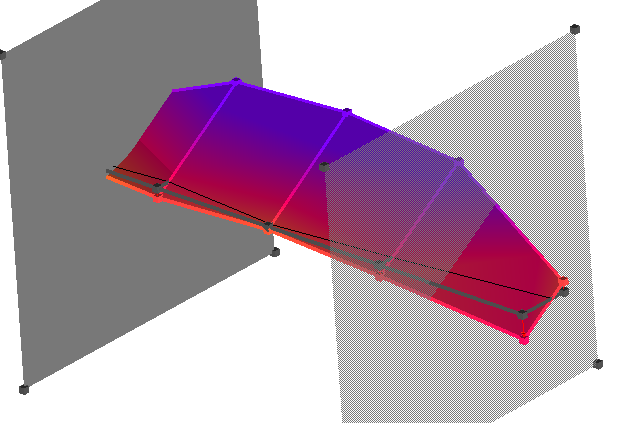}
        \caption{"Heave" mode (52~Hz)}
    \end{subfigure}
    \hfill
    \begin{subfigure}{0.32\textwidth}
        \includegraphics[width=\linewidth]{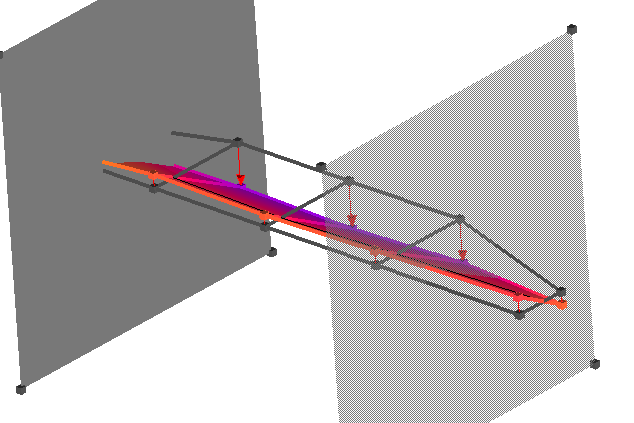}
        \caption{Lock-in coupling mode (91~Hz)}
    \end{subfigure}
    \hfill
    \begin{subfigure}{0.32\textwidth}
        \includegraphics[width=\linewidth]{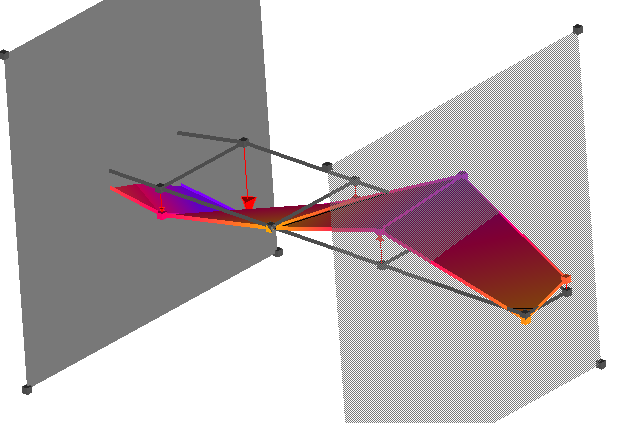}
        \caption{Harmonic of lock-in mode (182~Hz)}
    \end{subfigure}

    \vspace{1em}

    \textbf{(D) After lock-in}
    \label{fig:after_lockin}\\[0.3em]
    \begin{subfigure}{0.32\textwidth}
        \includegraphics[width=\linewidth]{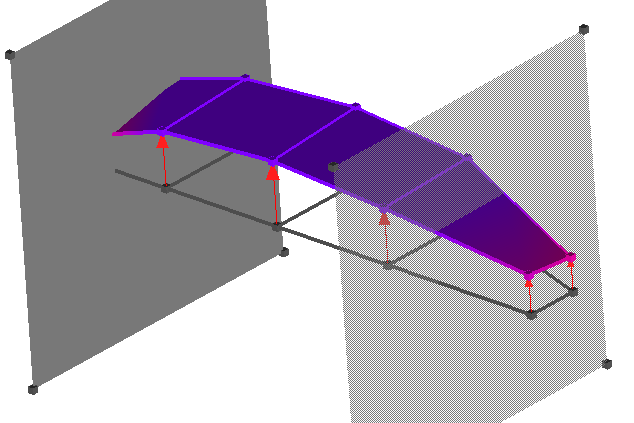}
        \caption{Heave mode (52~Hz)}
    \end{subfigure}
    \begin{subfigure}{0.32\textwidth}
        \includegraphics[width=\linewidth]{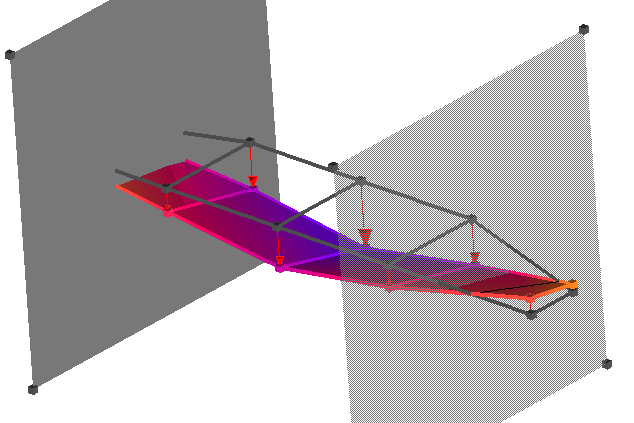}
        \caption{Bending mode (96~Hz)}
    \end{subfigure}

    \caption{Mode shapes identified via modal analysis at different stages of the fluid-structure interaction. Stages A, B, C and D are visible in Fig.~\ref{fig:Spectrograms}. Note: Mode shapes are normalized to unit  maximum displacement, thus amplitudes are not comparable between subfigures. 2 DOF configuration, $\alpha_0 = 3.5^\circ$, pitch DOF stiffness: $L_p = 100~\mathrm{mm}$, heave DOF stiffness: 3.}
    \label{fig:modes_shapes}
\end{figure*}

Modal identification analyses based on accelerometer data were performed, and the identified modal parameters are presented in Fig.~\ref{fig:modes_shapes}.
The black lines represent the undeformed sensor mesh, while the colored surfaces illustrate the modal displacement shapes associated with the spectrogram peaks shown in Fig.~\ref{fig:Spectrograms}.
The grey opaque and transparent rectangles represent the wind tunnel walls.
Four accelerometers were mounted at the wing roots (two on each side), outside the test section.
Mode shapes are normalized to unit maximum displacement.
Consequently, relative amplitudes of the mode shapes are not directly comparable.
The focus is on mode shape geometry and frequency evolution rather than absolute response levels.
Nevertheless, relative amplitude information can be inferred from peak magnitudes in the spectrograms (Fig.~\ref{fig:Spectrograms}).

Combining information from both Fig.~\ref{fig:Spectrograms} and Fig.~\ref{fig:modes_shapes}, the evolution of the flow-structure interaction can be divided into four transient stages, labeled A through D.

\textbf{Stage A — Pre-buffet (baseline structural dynamics).}
Prior to reaching the buffet threshold, the unsteady-pressure signal exhibits only a weak broadband response (Fig.~\ref{fig:Spectrogram_kulite_comments}).
In the structural spectrogram (Fig.~\ref{fig:Spectrogram_accelero}), the three natural frequencies of the system are clearly visible and well separated: the heave mode near 52~Hz, the out-of-plane bending mode near 96~Hz, and the pitch mode around 125~Hz.
This stage establishes the reference dynamics prior to any fluid-structure coupling.

\textbf{Stage B — Buffet development (emergence of a flow-induced mode).}
As the Mach number increases into the buffet regime, a distinct spectral peak appears near 70~Hz with a very slight second harmonic around 140~Hz.
This buffet frequency drifts slightly upward with Mach number.
Meanwhile, the pitch natural frequency decreases progressively, and the bending frequency shifts marginally.
A small increase in the heave-response level hints at weak energy exchange with the buffet mode, although part of this response may be simply forced excitation.
Modal analyses (Fig.~\ref{fig:modes_shapes}) show that the structural modes from Stage A persist, but a new mode emerges at the buffet frequency, involving a combination of pitch and bending motion.
The pitch mode cannot be cleanly isolated during this stage -- its frequency migrates and its amplitude remains low.
However, continuous inspection of the modal deformation across the 110--125~Hz region using the acquisition software revealed a consistently pitch-dominated shape, even though no distinct natural mode peak was identified in this range.

\textbf{Stage C — Lock-in (coalescence of buffet, bending, and pitch modes).}
Lock-in becomes evident when the buffet peak, the bending mode ($\approx95$~Hz), and the shifting pitch mode converge toward a common frequency around 91~Hz.
The dynamic response amplifies sharply, and the first harmonic of the coupled mode also attains large magnitude.
Although the lock-in frequency coincides with the bending mode frequency, the mode shape analysis reveals that the coupled phenomenon is predominantly driven by pitch motion.
The mode shape corresponding to the coalescence of the buffet-bending-pitch lock-in phenomenon is clearly dominated by pitch motion with only a slight bending contribution, which strongly suggests that the buffet-pitch interaction constitutes the core coupling mechanism.
The heave mode at 52~Hz also evolves: its deformation acquires a pronounced pitch-like character, showing that the lock-in mechanism has become the dominant driver of the structural response.
A harmonic of the coupled mode, near 182~Hz, exhibits combined pitch and second-bending characteristics.
This stage marks the full onset of fluid-structure coupling.

\textbf{Stage D — Post-lock-in (decay of coupling and partial recovery).}
When the Mach number is increased further, the coupled peak disappears and the structural modes re-emerge at their nominal frequencies.
However, the out-of-plane bending mode retains a markedly amplified response over a broader frequency range compared with Stage A.
The pitch mode, by contrast, is no longer distinctly identifiable in the power spectrum.
However, mode shapes near 120~Hz still exhibit pitch-type deformation.
This indicates that, although lock-in and buffet cease, the structural response does not revert to its initial characteristics and that vestiges of the coupling might persist.
Complementary observations in~\cite{sicard2025experimental} also revealed a hysteresis-like behavior when decreasing Mach number instead of increasing it during Stage D, maintaining the coupled state below its onset threshold -- a feature consistent with the strong nonlinear character of buffet-induced lock-in.

Modal evolution confirms that lock-in results from the gradual coalescence of the buffet, bending, and pitch modes into a single coupled response.
Prior to coupling, the bending frequency drifts slightly, while the buffet mode shifts upward more markedly.
The concurrent intensification of the buffet spectral peak indicates the rapid amplification of the dynamic response as coupling develops.
The subsequent convergence of the buffet and pitch frequencies toward the bending mode marks the final stage of the lock-in process.
Ultimately, the coupled frequency stabilizes slightly below the natural bending frequency, indicating that the bending mode constrains but does not fully dictate the dynamics, while pitch motion remains the dominant component of the coupled deformation.

\section{Conclusion}

An experimental test campaign was conducted at the ONERA S3Ch transonic tunnel, systematically investigating buffet-induced aeroelastic coupling across configurations of an OAT15A airfoil with 0, 1, and 2 structural DOFs in pitch and heave.
The results provide a comprehensive understanding of how structural flexibility governs the onset and development of buffet and lock-in under transonic buffet conditions.

In the fixed (0-DOF) configuration, the flow exhibited either fully developed buffet or pre-buffet transitional unsteadiness for Mach numbers between 0.69 and 0.71 and angles of attack from 3.25° to 4°.
Within this range, the unsteady response intensified progressively, with buffet frequencies ranging between 65 and 75~Hz, consistent with previous investigations, and serving as a reference for the subsequent aeroelastic configurations.
Some discrepancies with earlier wind tunnel tests at ONERA S3Ch (\cite{jacquin2009experimental}) are attributed to differences in wall-interference corrections and in the structural boundary conditions of the test setup.
These aspects must therefore be carefully accounted for in numerical model calibration efforts.
Complementary investigations are underway to assess and quantify the influence of wall-correction methodologies on the observed buffet characteristics.

Introducing a single pitch degree of freedom revealed that lock-in arises when the pitch natural frequency approaches the buffet frequency, while remaining above it.
Lock-in consistently occurred near Mach~0.70--0.71 and intensified as the angle of attack was increased.
The aeroelastic response of the model also increased while reducing pitch stiffness -- overall decreasing the ratio $f_\text{pitch}/f_\text{buffet}$ from approximately 2.2 to 1.5 over the campaign -- confirming the strong sensitivity of the coupled dynamics to frequency proximity, with amplitudes reaching levels comparable to limit-cycle oscillations for the ratio 1.5.
These findings are consistent with numerical predictions by \cite{giannelis2016dynamic}, where the ratio $f_\text{pitch}/f_\text{buffet} = 1.5$ represents the maximum amplitude of the interaction, and complement the experimental results of \cite{korthauer2023effect} on lower $f_\text{pitch}/f_\text{buffet}$ ratios (of 0.9, 1.0, and 1.2), on the same OAT15A airfoil geometry, where the ratio 1.2 demonstrated a similar lock-in phenomenon.
The observed upward shift of the buffet frequency toward structural modes further suggests that the structure actively drives the coupling in what Korthäuer described as structural-mode flutter.

Conversely, in the single-heave configuration, where the heave frequency lies below the buffet frequency, no lock-in was observed, despite the presence of an out-of-plane bending mode located just above the buffet frequency.
This bending mode, effectively behaving as a heave-like motion for a two-dimensional wing, did not induce coupling on its own.
The absence of lock-in is attributed to the large separation between the pitch and buffet frequencies ($f_\text{pitch}/f_\text{buffet} > 2$), confirming that proximity between the pitch and buffet modes is a necessary condition for lock-in to occur.

The two-DOF configuration clarified the coupling mechanism underlying buffet-induced lock-in.
Time-frequency analyses of unsteady pressure and accelerometer signals showed that, once buffet develops, a small increase in Mach number drives the buffet and pitch frequencies toward the natural bending frequency, producing a three-way fluid-structure interaction.
Modal analyses revealed that, during lock-in, the buffet, bending, and pitch modes merge into a single hybrid mode, exhibiting mixed pitch-bending characteristics.
The coupled frequency nearly coincides with the bending natural frequency, indicating near-resonance, yet remains slightly lower -- demonstrating that the bending mode modulates, but does not fully dictate, the coupled dynamics.
Furthermore, hysteresis-like behavior was observed during Mach number reduction (\cite{sicard2025experimental}), with the coupled state persisting below the onset threshold, consistent with the strong nonlinear character of buffet-induced lock-in.

Taken together, the results confirm that lock-in occurs only when the pitch mode lies sufficiently close to the buffet frequency.
Adding a heave degree of freedom below the buffet frequency does not fundamentally alter the phenomenology, whereas the out-of-plane bending mode above the buffet frequency plays a secondary but influential role in shaping the coupled response.

Future investigations employing high-fidelity numerical simulations that explicitly include the structural bending mode would further clarify the role of this mode in the aeroelastic dynamics of buffet-induced lock-in.

\bmhead{Data availability}
The data that support the findings of this study are available from the corresponding author upon reasonable request.

\bmhead{Acknowledgments}

Financial support in the frame of the Holistic Optical Metrology
for Aero-Elastic Research project from the European Union's Horizon
2020 research and innovation program under grant agreement
No.~769237 is gratefully acknowledged.
The authors would like to thank Jean-Charles Abart, Yannick Amossé, Florence Bouvier, Caroline Goudeau and Thomas Lambert for their commitment and support during the test campaign.

\bmhead{CRediT authorship contribution statement}

\textbf{Cédric Thémiot:} Conceptualization, Methodology, Investigation, Validation, Software, Formal Analysis, Data Curation, Visualization, Writing - Original Draft. Cédric Thémiot led the design phase of the wing model and revised test setup, conducted laboratory and wind tunnel tests, performed post-processing, and wrote the manuscript.
\textbf{Vincent Brion:} Conceptualization, Methodology, Supervision, Writing - Review \& Editing. Vincent Brion provided daily guidance, aerodynamic expertise, and reviewed the manuscript.
\textbf{Arnaud Lepage:} Conceptualization, Methodology, Supervision, Resources, Writing - Review \& Editing. Arnaud Lepage provided the original experimental test setup, daily guidance, aeroelastic expertise, and reviewed the manuscript.



\bibliography{sn-bibliography}


\begin{thebibliography}{34}
\ifx \bisbn   \undefined \def \bisbn  #1{ISBN #1}\fi
\ifx \binits  \undefined \def \binits#1{#1}\fi
\ifx \bauthor  \undefined \def \bauthor#1{#1}\fi
\ifx \batitle  \undefined \def \batitle#1{#1}\fi
\ifx \bjtitle  \undefined \def \bjtitle#1{#1}\fi
\ifx \bvolume  \undefined \def \bvolume#1{\textbf{#1}}\fi
\ifx \byear  \undefined \def \byear#1{#1}\fi
\ifx \bissue  \undefined \def \bissue#1{#1}\fi
\ifx \bfpage  \undefined \def \bfpage#1{#1}\fi
\ifx \blpage  \undefined \def \blpage #1{#1}\fi
\ifx \burl  \undefined \def \burl#1{\textsf{#1}}\fi
\ifx \doiurl  \undefined \def \doiurl#1{\url{https://doi.org/#1}}\fi
\ifx \betal  \undefined \def \betal{\textit{et al.}}\fi
\ifx \binstitute  \undefined \def \binstitute#1{#1}\fi
\ifx \binstitutionaled  \undefined \def \binstitutionaled#1{#1}\fi
\ifx \bctitle  \undefined \def \bctitle#1{#1}\fi
\ifx \beditor  \undefined \def \beditor#1{#1}\fi
\ifx \bpublisher  \undefined \def \bpublisher#1{#1}\fi
\ifx \bbtitle  \undefined \def \bbtitle#1{#1}\fi
\ifx \bedition  \undefined \def \bedition#1{#1}\fi
\ifx \bseriesno  \undefined \def \bseriesno#1{#1}\fi
\ifx \blocation  \undefined \def \blocation#1{#1}\fi
\ifx \bsertitle  \undefined \def \bsertitle#1{#1}\fi
\ifx \bsnm \undefined \def \bsnm#1{#1}\fi
\ifx \bsuffix \undefined \def \bsuffix#1{#1}\fi
\ifx \bparticle \undefined \def \bparticle#1{#1}\fi
\ifx \barticle \undefined \def \barticle#1{#1}\fi
\bibcommenthead
\ifx \bconfdate \undefined \def \bconfdate #1{#1}\fi
\ifx \botherref \undefined \def \botherref #1{#1}\fi
\ifx \url \undefined \def \url#1{\textsf{#1}}\fi
\ifx \bchapter \undefined \def \bchapter#1{#1}\fi
\ifx \bbook \undefined \def \bbook#1{#1}\fi
\ifx \bcomment \undefined \def \bcomment#1{#1}\fi
\ifx \oauthor \undefined \def \oauthor#1{#1}\fi
\ifx \citeauthoryear \undefined \def \citeauthoryear#1{#1}\fi
\ifx \endbibitem  \undefined \def \endbibitem {}\fi
\ifx \bconflocation  \undefined \def \bconflocation#1{#1}\fi
\ifx \arxivurl  \undefined \def \arxivurl#1{\textsf{#1}}\fi
\csname PreBibitemsHook\endcsname

\bibitem[\protect\citeauthoryear{Accorinti
  et~al.}{2022}]{accorinti2022experimental}
\begin{barticle}
\bauthor{\bsnm{Accorinti}, \binits{A.}},
\bauthor{\bsnm{Baur}, \binits{T.}},
\bauthor{\bsnm{Scharnowski}, \binits{S.}},
\bauthor{\bsnm{K{\"a}hler}, \binits{C.J.}}:
\batitle{Experimental investigation of transonic shock buffet on an oat15a
  profile}.
\bjtitle{AIAA Journal}
\bvolume{60}(\bissue{11}),
\bfpage{6289}--\blpage{6300}
(\byear{2022})
\end{barticle}
\endbibitem

\bibitem[\protect\citeauthoryear{Accorinti
  et~al.}{2023}]{accorinti2023characterization}
\begin{barticle}
\bauthor{\bsnm{Accorinti}, \binits{A.}},
\bauthor{\bsnm{Korth{\"a}uer}, \binits{T.}},
\bauthor{\bsnm{Scharnowski}, \binits{S.}},
\bauthor{\bsnm{K{\"a}hler}, \binits{C.J.}}:
\batitle{Characterization of transonic shock oscillations over the span of an
  oat15a profile}.
\bjtitle{Experiments in Fluids}
\bvolume{64}(\bissue{3}),
\bfpage{61}
(\byear{2023})
\end{barticle}
\endbibitem

\bibitem[\protect\citeauthoryear{Albano and Rodden}{1969}]{albano1969doublet}
\begin{barticle}
\bauthor{\bsnm{Albano}, \binits{E.}},
\bauthor{\bsnm{Rodden}, \binits{W.P.}}:
\batitle{A doublet-lattice method for calculating lift distributions on
  oscillating surfaces in subsonic flows.}
\bjtitle{AIAA journal}
\bvolume{7}(\bissue{2}),
\bfpage{279}--\blpage{285}
(\byear{1969})
\end{barticle}
\endbibitem

\bibitem[\protect\citeauthoryear{Blair}{1992}]{blair1992compilation}
\begin{bbook}
\bauthor{\bsnm{Blair}, \binits{M.}}:
\bbtitle{A Compilation of the Mathematics Leading to the Doublet-lattice
  Method}
vol. \bseriesno{92}.
\bpublisher{Fliight Dynamics Directorate, Air Force Wright Laboratory, Air
  Force Systems~…}, \blocation{???}
(\byear{1992})
\end{bbook}
\endbibitem

\bibitem[\protect\citeauthoryear{Crouch et~al.}{2009}]{crouch2009origin}
\begin{barticle}
\bauthor{\bsnm{Crouch}, \binits{J.D.}},
\bauthor{\bsnm{Garbaruk}, \binits{A.}},
\bauthor{\bsnm{Magidov}, \binits{D.}},
\bauthor{\bsnm{Travin}, \binits{A.}}:
\batitle{Origin of transonic buffet on aerofoils}.
\bjtitle{Journal of fluid mechanics}
\bvolume{628},
\bfpage{357}--\blpage{369}
(\byear{2009})
\end{barticle}
\endbibitem

\bibitem[\protect\citeauthoryear{Giannelis
  et~al.}{2018}]{giannelis2018influence}
\begin{botherref}
\oauthor{\bsnm{Giannelis}, \binits{N.}},
\oauthor{\bsnm{Levinski}, \binits{O.}},
\oauthor{\bsnm{Vio}, \binits{G.}}:
Influence of mach number and angle of attack on the two-dimensional transonic
  buffet phenomenon.
Aerospace Science and Technology
\textbf{78}
(2018)
\end{botherref}
\endbibitem

\bibitem[\protect\citeauthoryear{Giannelis and
  Vio}{2016}]{giannelis2016aeroelastic}
\begin{bchapter}
\bauthor{\bsnm{Giannelis}, \binits{N.}},
\bauthor{\bsnm{Vio}, \binits{G.}}:
\bctitle{Aeroelastic interactions of a supercritical aerofoil in the presence
  of transonic shock buffet}.
In: \bbtitle{Proceedings of the 54th AIAA Aerospace Sciences Meeting, San
  Diego, CA}
(\byear{2016})
\end{bchapter}
\endbibitem

\bibitem[\protect\citeauthoryear{Giannelis and
  Vio}{2017}]{giannelis2017investigation}
\begin{bchapter}
\bauthor{\bsnm{Giannelis}, \binits{N.F.}},
\bauthor{\bsnm{Vio}, \binits{G.A.}}:
\bctitle{Investigation of frequency lock-in phenomena on a supercritical
  aerofoil in the presence of transonic shock oscillations}.
In: \bbtitle{Proceedings of the 17th International Forum on Aeroelasticity and
  Structural Dynamics, Como, Italy},
pp. \bfpage{25}--\blpage{28}
(\byear{2017})
\end{bchapter}
\endbibitem

\bibitem[\protect\citeauthoryear{Giannelis et~al.}{2016}]{giannelis2016dynamic}
\begin{bchapter}
\bauthor{\bsnm{Giannelis}, \binits{N.}},
\bauthor{\bsnm{Vio}, \binits{G.}},
\bauthor{\bsnm{Dimitriadis}, \binits{G.}}:
\bctitle{Dynamic interactions of a supercritical aerofoil in the presence of
  transonic shock buffet}.
In: \bbtitle{Proceedings of the 27th International Conference on Noise and
  Vibration Engineering, Leuven, Belgium}
(\byear{2016})
\end{bchapter}
\endbibitem

\bibitem[\protect\citeauthoryear{Giannelis et~al.}{2017}]{giannelis2017review}
\begin{barticle}
\bauthor{\bsnm{Giannelis}, \binits{N.F.}},
\bauthor{\bsnm{Vio}, \binits{G.A.}},
\bauthor{\bsnm{Levinski}, \binits{O.}}:
\batitle{A review of recent developments in the understanding of transonic
  shock buffet}.
\bjtitle{Progress in Aerospace Sciences}
\bvolume{92},
\bfpage{39}--\blpage{84}
(\byear{2017})
\end{barticle}
\endbibitem

\bibitem[\protect\citeauthoryear{Gao and Zhang}{2020}]{gao2020transonic}
\begin{barticle}
\bauthor{\bsnm{Gao}, \binits{C.}},
\bauthor{\bsnm{Zhang}, \binits{W.}}:
\batitle{Transonic aeroelasticity: A new perspective from the fluid mode}.
\bjtitle{Progress in Aerospace Sciences}
\bvolume{113},
\bfpage{100596}
(\byear{2020})
\end{barticle}
\endbibitem

\bibitem[\protect\citeauthoryear{Hilton and
  Fowler}{1947}]{hilton1947photographs}
\begin{botherref}
\oauthor{\bsnm{Hilton}, \binits{W.F.}},
\oauthor{\bsnm{Fowler}, \binits{R.}}:
Photographs of shock wave movement
(1947)
\end{botherref}
\endbibitem

\bibitem[\protect\citeauthoryear{Hartmann et~al.}{2013}]{hartmann2013coupled}
\begin{barticle}
\bauthor{\bsnm{Hartmann}, \binits{A.}},
\bauthor{\bsnm{Klaas}, \binits{M.}},
\bauthor{\bsnm{Schr{\"o}der}, \binits{W.}}:
\batitle{Coupled airfoil heave/pitch oscillations at buffet flow}.
\bjtitle{AIAA journal}
\bvolume{51}(\bissue{7}),
\bfpage{1542}--\blpage{1552}
(\byear{2013})
\end{barticle}
\endbibitem

\bibitem[\protect\citeauthoryear{Iovnovich and
  Raveh}{2012}]{iovnovich2012transonic}
\begin{barticle}
\bauthor{\bsnm{Iovnovich}, \binits{M.}},
\bauthor{\bsnm{Raveh}, \binits{D.}}:
\batitle{Transonic unsteady aerodynamics in the vicinity of shock-buffet
  instability}.
\bjtitle{Journal of Fluids and Structures}
\bvolume{29},
\bfpage{131}--\blpage{142}
(\byear{2012})
\end{barticle}
\endbibitem

\bibitem[\protect\citeauthoryear{Jacquin et~al.}{2016}]{jacquin2016testing}
\begin{botherref}
\oauthor{\bsnm{Jacquin}, \binits{L.}},
\oauthor{\bsnm{Brion}, \binits{V.}},
\oauthor{\bsnm{Molton}, \binits{P.}},
\oauthor{\bsnm{Sipp}, \binits{D.}},
\oauthor{\bsnm{Dandois}, \binits{J.}},
\oauthor{\bsnm{Deck}, \binits{S.}},
\oauthor{\bsnm{Sartor}, \binits{F.}},
\oauthor{\bsnm{Coustols}, \binits{E.}},
\oauthor{\bsnm{Caruana}, \binits{D.}}:
Testing in aerodynamics research at onera: The example of the transonic buffet.
Aerospace Lab
(12),
1
(2016)
\end{botherref}
\endbibitem

\bibitem[\protect\citeauthoryear{Jacquin
  et~al.}{2009}]{jacquin2009experimental}
\begin{barticle}
\bauthor{\bsnm{Jacquin}, \binits{L.}},
\bauthor{\bsnm{Molton}, \binits{P.}},
\bauthor{\bsnm{Deck}, \binits{S.}},
\bauthor{\bsnm{Maury}, \binits{B.}},
\bauthor{\bsnm{Soulevant}, \binits{D.}}:
\batitle{Experimental study of shock oscillation over a transonic supercritical
  profile}.
\bjtitle{AIAA journal}
\bvolume{47}(\bissue{9}),
\bfpage{1985}--\blpage{1994}
(\byear{2009})
\end{barticle}
\endbibitem

\bibitem[\protect\citeauthoryear{Korth{\"a}uer
  et~al.}{2023a}]{korthauer2023effect}
\begin{barticle}
\bauthor{\bsnm{Korth{\"a}uer}, \binits{T.}},
\bauthor{\bsnm{Accorinti}, \binits{A.}},
\bauthor{\bsnm{Scharnowski}, \binits{S.}},
\bauthor{\bsnm{K{\"a}hler}, \binits{C.J.}}:
\batitle{Effect of mach number and pitching eigenfrequency on transonic buffet
  onset}.
\bjtitle{AIAA Journal}
\bvolume{61}(\bissue{1}),
\bfpage{112}--\blpage{124}
(\byear{2023})
\end{barticle}
\endbibitem

\bibitem[\protect\citeauthoryear{Korth{\"a}uer
  et~al.}{2023b}]{korthauer2023experimental}
\begin{barticle}
\bauthor{\bsnm{Korth{\"a}uer}, \binits{T.}},
\bauthor{\bsnm{Accorinti}, \binits{A.}},
\bauthor{\bsnm{Scharnowski}, \binits{S.}},
\bauthor{\bsnm{K{\"a}hler}, \binits{C.J.}}:
\batitle{Experimental investigation of transonic buffeting, frequency lock-in
  and their dependence on structural characteristics}.
\bjtitle{Journal of Fluids and Structures}
\bvolume{122},
\bfpage{103975}
(\byear{2023})
\end{barticle}
\endbibitem

\bibitem[\protect\citeauthoryear{Kokmanian
  et~al.}{2022}]{kokmanian2022investigating}
\begin{barticle}
\bauthor{\bsnm{Kokmanian}, \binits{K.}},
\bauthor{\bsnm{Scharnowski}, \binits{S.}},
\bauthor{\bsnm{Sch{\"a}fer}, \binits{C.}},
\bauthor{\bsnm{Accorinti}, \binits{A.}},
\bauthor{\bsnm{Baur}, \binits{T.}},
\bauthor{\bsnm{K{\"a}hler}, \binits{C.J.}}:
\batitle{Investigating the flow field dynamics of transonic shock buffet using
  particle image velocimetry}.
\bjtitle{Experiments in Fluids}
\bvolume{63}(\bissue{9}),
\bfpage{149}
(\byear{2022})
\end{barticle}
\endbibitem

\bibitem[\protect\citeauthoryear{Lepage et~al.}{2015}]{lepage2015complete}
\begin{bchapter}
\bauthor{\bsnm{Lepage}, \binits{A.}},
\bauthor{\bsnm{Amosse}, \binits{Y.}},
\bauthor{\bsnm{Le~Bihan}, \binits{D.}},
\bauthor{\bsnm{Poussot-Vassal}, \binits{C.}},
\bauthor{\bsnm{Brion}, \binits{V.}},
\bauthor{\bsnm{Rantet}, \binits{E.}}:
\bctitle{A complete experimental investigation of gust load: from generation to
  active control}.
In: \bbtitle{International Forum on Aeroelasticity and Structural Dynamics.
  Saint Petersburg, Russia}
(\byear{2015})
\end{bchapter}
\endbibitem

\bibitem[\protect\citeauthoryear{Lee}{2001}]{lee2001self}
\begin{barticle}
\bauthor{\bsnm{Lee}, \binits{B.}}:
\batitle{Self-sustained shock oscillations on airfoils at transonic speeds}.
\bjtitle{Progress in Aerospace Sciences}
\bvolume{37}(\bissue{2}),
\bfpage{147}--\blpage{196}
(\byear{2001})
\end{barticle}
\endbibitem

\bibitem[\protect\citeauthoryear{McDevitt and Okuno}{1985}]{mcdevitt1985static}
\begin{botherref}
\oauthor{\bsnm{McDevitt}, \binits{J.B.}},
\oauthor{\bsnm{Okuno}, \binits{A.F.}}:
Static and dynamic pressure measurements on a naca 0012 airfoil in the ames
  high reynolds number facility.
Technical report
(1985)
\end{botherref}
\endbibitem

\bibitem[\protect\citeauthoryear{Nitzsche}{2009}]{nitzsche2009numerical}
\begin{botherref}
\oauthor{\bsnm{Nitzsche}, \binits{J.}}:
A numerical study on aerodynamic resonance in transonic separated flow
(2009)
\end{botherref}
\endbibitem

\bibitem[\protect\citeauthoryear{Nitzsche et~al.}{2019}]{nitzsche2019fluid}
\begin{bchapter}
\bauthor{\bsnm{Nitzsche}, \binits{J.}},
\bauthor{\bsnm{Ringel}, \binits{L.M.}},
\bauthor{\bsnm{Kaiser}, \binits{C.}},
\bauthor{\bsnm{Hennings}, \binits{H.}}:
\bctitle{Fluid-mode flutter in plane transonic flows}.
In: \bbtitle{International Forum on Aeroelasticity and Structural Dynamics
  2019, IFASD 2019}
(\byear{2019})
\end{bchapter}
\endbibitem

\bibitem[\protect\citeauthoryear{Pearcey}{1955}]{pearcey1955some}
\begin{botherref}
\oauthor{\bsnm{Pearcey}, \binits{H.}}:
Some effects of shock-induced separation of turbulent boundary layers in
  transonic flow past aerofoils
(1955)
\end{botherref}
\endbibitem

\bibitem[\protect\citeauthoryear{Pearcey et~al.}{1968}]{pearcey1968interaction}
\begin{bbook}
\bauthor{\bsnm{Pearcey}, \binits{H.}},
\bauthor{\bsnm{Haines}, \binits{A.}},
\bauthor{\bsnm{Osborne}, \binits{J.}}:
\bbtitle{The Interaction Between Local Effects at the Shock and Rear
  Separation-a Source of Significant Scale Effects in Wind-tunnel Tests on
  Aerofoils and Wings}.
\bpublisher{NPL}, \blocation{???}
(\byear{1968})
\end{bbook}
\endbibitem

\bibitem[\protect\citeauthoryear{Raveh}{2009}]{raveh2009numerical}
\begin{barticle}
\bauthor{\bsnm{Raveh}, \binits{D.E.}}:
\batitle{Numerical study of an oscillating airfoil in transonic buffeting
  flows}.
\bjtitle{AIAA journal}
\bvolume{47}(\bissue{3}),
\bfpage{505}--\blpage{515}
(\byear{2009})
\end{barticle}
\endbibitem

\bibitem[\protect\citeauthoryear{Raveh and Dowell}{2011}]{raveh2011frequency}
\begin{barticle}
\bauthor{\bsnm{Raveh}, \binits{D.}},
\bauthor{\bsnm{Dowell}, \binits{E.}}:
\batitle{Frequency lock-in phenomenon for oscillating airfoils in buffeting
  flows}.
\bjtitle{Journal of Fluids and Structures}
\bvolume{27}(\bissue{1}),
\bfpage{89}--\blpage{104}
(\byear{2011})
\end{barticle}
\endbibitem

\bibitem[\protect\citeauthoryear{Raveh and Dowell}{2014}]{raveh2014aeroelastic}
\begin{barticle}
\bauthor{\bsnm{Raveh}, \binits{D.E.}},
\bauthor{\bsnm{Dowell}, \binits{E.H.}}:
\batitle{Aeroelastic responses of elastically suspended airfoil systems in
  transonic buffeting flows}.
\bjtitle{AIAA journal}
\bvolume{52}(\bissue{5}),
\bfpage{926}--\blpage{934}
(\byear{2014})
\end{barticle}
\endbibitem

\bibitem[\protect\citeauthoryear{Richard and
  Harrison}{1979}]{richard1979program}
\begin{botherref}
\oauthor{\bsnm{Richard}, \binits{M.}},
\oauthor{\bsnm{Harrison}, \binits{B.}}:
A program to compute three-dimensional subsonic unsteady aerodynamic
  characteristics using the doublet lattice method, l216 (dubflx). volume 1:
  Engineering and usage.
Technical report,
NASA
(1979)
\end{botherref}
\endbibitem

\bibitem[\protect\citeauthoryear{Scharnowski
  et~al.}{2022}]{scharnowski2022shock}
\begin{barticle}
\bauthor{\bsnm{Scharnowski}, \binits{S.}},
\bauthor{\bsnm{Kokmanian}, \binits{K.}},
\bauthor{\bsnm{Sch{\"a}fer}, \binits{C.}},
\bauthor{\bsnm{Baur}, \binits{T.}},
\bauthor{\bsnm{Accorinti}, \binits{A.}},
\bauthor{\bsnm{K{\"a}hler}, \binits{C.J.}}:
\batitle{Shock-buffet analysis on a supercritical airfoil with a pitching
  degree of freedom}.
\bjtitle{Experiments in Fluids}
\bvolume{63}(\bissue{6}),
\bfpage{1}--\blpage{13}
(\byear{2022})
\end{barticle}
\endbibitem

\bibitem[\protect\citeauthoryear{Sicard et~al.}{2025}]{sicard2025experimental}
\begin{bchapter}
\bauthor{\bsnm{Sicard}, \binits{L.}},
\bauthor{\bsnm{Th{\'e}miot}, \binits{C.}},
\bauthor{\bsnm{Brion}, \binits{V.}}:
\bctitle{Experimental characterization of the transonic buffet lock-in
  phenomenon for an airfoil in free pitching and plunging motion}.
In: \bbtitle{AERO 2025-59th 3AF International Conference on Applied
  Aerodynamics}
(\byear{2025})
\end{bchapter}
\endbibitem

\bibitem[\protect\citeauthoryear{Th{\'e}miot et~al.}{2023}]{themiot2023wind}
\begin{bchapter}
\bauthor{\bsnm{Th{\'e}miot}, \binits{C.}},
\bauthor{\bsnm{Brion}, \binits{V.}},
\bauthor{\bsnm{Lepage}, \binits{A.}}:
\bctitle{Wind tunnel experiment on a pitch and plunge free airfoil under
  transonic buffet}.
In: \bbtitle{AERO 2023: 57th 3AF International Conference on Applied
  Aerodynamics}
(\byear{2023})
\end{bchapter}
\endbibitem

\bibitem[\protect\citeauthoryear{Tijdeman}{1977}]{tijdeman1977investigations}
\begin{botherref}
\oauthor{\bsnm{Tijdeman}, \binits{H.}}:
Investigations of the transonic flow around oscillating airfoils.
NLR-TR 77090 U
(1977)
\end{botherref}
\endbibitem

\end{thebibliography}


\end{document}